\documentclass[aps,superscriptaddress,twocolumn,amsmath,amssymb]{revtex4-1}

\usepackage{mathtools}
\usepackage{mathrsfs}
\usepackage{amsmath}
\usepackage[mathscr]{eucal}
\usepackage{xfrac}
\usepackage{graphicx}
\usepackage{dcolumn}
\usepackage{bm}
\usepackage{color}
\usepackage{tabularx}
\usepackage[colorlinks,linkcolor=OrangeRed,citecolor=RoyalBlue,urlcolor=NavyBlue]{hyperref}
\usepackage[normalem]{ulem}
\usepackage[all]{hypcap}
\usepackage[dvipsnames,usenames,table]{xcolor}

\newcommand{\nocontentsline}[3]{}
\newcommand{\tocless}[2]{\bgroup\let\addcontentsline=\nocontentsline#1{#2}\egroup}

\newcommand{\be}{\begin{equation}}
\newcommand{\ee}{\end{equation}}
\newcommand{\bk}{{\bf k}}
\newcommand{\bK}{{\bf K}}
\newcommand{\bq}{{\bf q}}

\newcommand{\br}{{\bf r}}

\newcommand{\bI}{{\bf I}}
\newcommand{\bs}{{\bf s}}
\newcommand{\bS}{{\bf S}}

\newcommand{\bn}{{\bf n}}

\newcommand{\Tm}{\mathcal{T}}
\newcommand{\Cm}{\mathcal{C}}
\newcommand{\Pm}{\mathcal{P}}
\newcommand{\trans}{\mathsf{T}}

\newcommand{\bdelta}{\boldsymbol{\Delta}}

\newcommand{\beq}{\begin{eqnarray}}
\newcommand{\eeq}{\end{eqnarray}}
\newcommand{\bea}{\begin{align}}
\newcommand{\eea}{\end{align}}
\newcommand{\beqq}{\begin{eqnarray*}}
\newcommand{\eeqq}{\end{eqnarray*}}

\newcommand{\makebf}[1]{\boldsymbol{#1} }

\newcommand{\bracket}[2]{\langle #1 | #2 \rangle }
\newcommand{\bra}[1]{\langle #1 | }
\newcommand{\ket}[1]{ | #1 \rangle }

\newcommand{\tr}[1]{\text{Tr}\,[#1 ]  }

\newcommand{\spinS}[2]{\mathsf{S}_{#1 #2}}
\newcommand{\spinJ}[2]{\mathsf{J}_{#1 #2}(\hat \bk)}

\newcommand{\up}{\uparrow}
\newcommand{\down}{\downarrow}

\begin{document}


\title{Pairing states of spin-$\frac{3}{2}$ fermions: Symmetry-enforced topological gap functions}

\author{J\"{o}rn W. F. Venderbos}
\email{jwfv@sas.upenn.edu}
\affiliation{Department of Physics, Massachusetts Institute of Technology, Cambridge, Massachusetts 02139, USA }%
\affiliation{The Makineni Theoretical Laboratories, Department of Chemistry, University of Pennsylvania, Philadelphia, Pennsylvania 19104, USA}%
\affiliation{Department of Physics and Astronomy, University of Pennsylvania, Philadelphia, Pennsylvania 19104, USA}%
\author{Lucile Savary}
\affiliation{Department of Physics, Massachusetts Institute of Technology, Cambridge, Massachusetts 02139, USA }%
\affiliation{Laboratoire de physique, CNRS, \'{E}cole Normale Sup\'{e}rieure de Lyon, 46, all\'{e}e d'Italie, 69007 Lyon}
\author{Jonathan Ruhman}
\affiliation{Department of Physics, Massachusetts Institute of Technology, Cambridge, Massachusetts 02139, USA }
\author{Patrick A. Lee}
\affiliation{Department of Physics, Massachusetts Institute of Technology, Cambridge, Massachusetts 02139, USA }
\author{Liang Fu}
\affiliation{Department of Physics, Massachusetts Institute of Technology, Cambridge, Massachusetts 02139, USA }

\date{\today}

\begin{abstract}
We study the topological properties of superconductors with paired $j=\frac{3}{2}$ quasiparticles. Higher spin Fermi surfaces can arise, for instance, in strongly spin-orbit coupled band-inverted semimetals. Examples include the Bi-based half-Heusler materials, which have recently been established as low-temperature and low-carrier density superconductors. Motivated by this experimental observation, we obtain a comprehensive symmetry-based classification of topological pairing states in systems with higher angular momentum Cooper pairing. Our study consists of two main parts. First, we develop the phenomenological theory of multicomponent (i.e., higher angular momentum) pairing by classifying the stationary points of the free energy within a Ginzburg-Landau framework. Based on the symmetry classification of stationary pairing states, we then derive the symmetry-imposed constraints on their gap structures. We find that, depending on the symmetry quantum numbers of the Cooper pairs, different types of topological pairing states can occur: fully gapped topological superconductors in class DIII, Dirac superconductors and superconductors hosting Majorana fermions. Notably, we find a series of nematic fully gapped topological superconductors, as well as double- and triple-Dirac superconductors, with quadratic and cubic dispersion, respectively. Our approach, applied here to the case of $j=\frac{3}{2}$ Cooper pairing, is rooted in the symmetry properties of pairing states, and can therefore also be applied to other systems with higher angular momentum and high-spin pairing. We conclude by relating our results to experimentally accessible signatures in thermodynamic and dynamic probes.
\end{abstract}

\maketitle


\section{Introduction}

In condensed matter physics, the study of superconductors has traditionally been guided by two defining characteristics of a bulk superconductor: the nature of the pairing order parameter and the mechanism of Cooper pairing \cite{sigrist91,carbotte90,scalapino12,chubukov13}. Recent years, however, have witnessed great progress in understanding phases of quantum matter from the perspective of topology. In particular, in the case of superconductors, it has become clear that a global property of the Cooper pair wavefunction, encoded in its topology, constitutes a third defining characteristic. Nontrivial topology leads to the presence of quasiparticle excitations on surfaces and edges \cite{schnyder08,kane10,qi11,mizushima16,sato17}. Specifically, the class of topological superconductors---in much the same way as topological insulators and topological semimetals---can be distinguished from ordinary superconductors by gapless quasiparticle excitations on the boundary, protected by the bulk superconducting gap structure. The latter is a manifestation of the bulk-boundary correspondence, which establishes an inherent link between surface properties and bulk topology. 

Topological superconductors with a bulk pairing gap are defined by a gap structure which cannot be adiabatically deformed into an $s$-wave superconductor without closing the pairing gap. Evidently, this implies that the question of pairing symmetry and bulk topology are intimately related. Indeed, time-reversal invariant topological superconductors in class DIII are known to require odd-parity pairing \cite{fu10,sato10}. The close connection between unconventional pairing symmetry and bulk topology is also manifest in topological nodal superconductors, i.e., superconductors with topologically protected nodal degeneracies in the bulk quasiparticle spectrum and distinctive gapless excitations at the surface \cite{matsuura13}. A famous example of the latter are the topological bulk point nodes of the superfluid $^3$He A-phase, which originate from the time-reversal breaking chiral pairing \cite{volovik}. Therefore, superconductors with unconventional pairing symmetry generally inspire the question whether they realize topological pairing states, and thus have protected Andreev surface states.

In this paper we address this question for superconductors with a semimetallic normal state characterized by quadratically dispersing spin-orbit coupled $j=\frac32$ bands. An important motivation for this undertaking is the observation of superconductivity in the class of Bi-based half-Heusler materials \emph{A}PtBi and \emph{A}PdBi, where \emph{A} can be a rare-earth element or Y/Lu \cite{chadov10,goll08,butch11,pan13,tafti13,xu14,nikitin15,nakajima15,liu16,zhang16,kim16}. Experimental evidence, in particular peneration depth measurements reported in Ref. \onlinecite{kim16}, has given indications that the pairing in YPtBi is unconventional. 

An additional incentive to consider the interplay of unconventional pairing and topology in spin-orbit coupled $j=\frac32$ bands is the possibility of high-spin Cooper pairing. This was recognized in important papers focusing on a specific set of fully gapped pairing states \cite{fang15,li16} and on-site pairings \cite{brydon16}. Reference \onlinecite{brydon16} in particular has set the stage for studying superconductivity in the half-Heusler compounds \cite{meinert16,boettcher16,agterberg17,savary17,boettcher17,timm17,yang17}; the present authors have investigated the pairing instabilities in the $p$-wave pairing channels \cite{savary17}. 

Whereas previous work has focussed primarily on the question of pairing symmetry, specifically in the context of materials such as YPtBi, the aim of this paper is to provide a comprehensive topological gap structure classification of spin $j=\frac32$ pairing states. Such classification, which encompasses all pairing channels, is desirable for the practical purpose of interpreting ongoing and future experiments, and stands to enable important progress in identifying the nature of the pairing order parameter in $j=\frac32$ systems. 

We proceed in two main steps. First, for multicomponent pairing channels, i.e., channels of Cooper pairing with nonzero total angular momentum, we obtain the stationary points of the free energy within a Ginzburg-Landau expansion using a symmetry-based strategy developed for $^3$He \cite{bruder86}. Since only pairing states corresponding to stationary points can be minima of the free energy, any analysis of gap structure can be limited to this set of possible superconducting ground states determined by energetics. This, in practice, is a significant simplification. The second step is then to systematically analyze the topology of gap structures of each stationary pairing state by deriving the constraints imposed on the superconducting gap function by discrete (e.g., time-reversal, inversion, mirror) and rotational symmetries.

Notably, we find a series of fully gapped topological superconductors which spontaneously break rotation symmetry and have a nematic axis \cite{fu14,venderbos16}. In addition, we obtain different classes of point nodal superconductors, hosting low-energy Dirac or Majorana bulk quasiparticles with dispersion relations which depend on topological properties of the point node. Importantly, despite starting from a normal state with full rotational symmetry (emergent at low-energies), our formalism includes pairing states with discrete spatial symmetry and can therefore be viewed as including crystal anisotropy effects. Furthermore, since our approach relies on symmetry arguments, the results of our work are relevant to a broad range of spin-orbit coupled systems with higher angular momentum pairing.


{\hypersetup{linkcolor=black}
\tableofcontents
}

\section{Multicomponent pairing of $j=\frac32$ quasiparticles \label{sec:pairing}}

\subsection{Electronic structure of quadratic semimetals \label{ssec:electronicstruct}}

We begin by introducing the Hamiltonian of the normal state electronic structure close to the semi-metallic touching point at the zone center. We assume that other electron or hole pockets are absent. The normal state Hamiltonian is expressed as 
\be
H_0 = \sum_\bk c_{\bk\alpha}^\dagger (h_\bk)_{\alpha\beta}c_{\bk\beta}, \label{eq:H0}
\ee
where $c_{\bk} = (c_{\bk \frac{3}{2}} ,c_{\bk\frac{1}{2}} , c_{\bk, -\frac{1}{2}} , c_{\bk,-\frac{3}{2}} )^\trans$ are the $j=\frac32$ quasiparticle annihilation operators and $h_\bk$ takes the isotropic Luttinger form \cite{luttinger56,murakami04}
\be
h_\bk= (\kappa_1+\frac{5}{4}\kappa_2) \frac{\bk^2}{2m}   -  \frac{\kappa_2}{2m} (\bk \cdot \bS )^2 -\mu . \label{eq:hk}
\ee
Here, $m$ is an effective mass, $\mu$ is the chemical potential, and $\bS = (S_x,S_y,S_z)^\trans$ are the three spin matrices. (Explicit expressions of the spin matrices are provided in Appendix \ref{app:spinmat}.)
The Luttinger Hamiltonian describes a touching of quadratically dispersing bands at $\Gamma$ which are spin-orbit split by the term $(\bk \cdot \bS )^2$. As a consequence of both time-reversal ($\Theta$) and inversion symmetry ($P$) the bands remain twofold degenerate at each momentum $\bk$. 

The Luttinger Hamiltonian can be diagonalized and brought into the form 
\be
H_0 = \sum_\bk \varepsilon^{v}_\bk f_{\bk}^\dagger f_{\bk} +\varepsilon^{c}_\bk d_{\bk}^\dagger d_{\bk}, \label{eq:H0diag}
\ee
where the energies measured with respect to the chemical potential are given by
\be
\varepsilon^{c,v}_\bk = (\kappa_1\pm\kappa_2)\frac{\bk^2}{2m} -\mu,    \label{eq:energies}
\ee
and the operators $f^\dagger_{\bk} = (f^\dagger_{\bk\up} ,f^\dagger_{\bk\down} )$ and $d^\dagger_{\bk} = (d^\dagger_{\bk\up} ,d^\dagger_{\bk\down} )$ create quasiparticles in the energy eigenstates. The twofold degeneracy of each band defines an effective pseudospin degree of freedom, which we denote by $\up,\down$. 
Equation \eqref{eq:energies} shows that the coefficients $\kappa_{1,2}$ directly relate to the band curvatures. In this work we will particularize to the regime where $\kappa_2> \kappa_1 > 0$. These conditions ensure that one pair of degenerate bands is electron-like and curving upward, the $\varepsilon^{c}_\bk$ solution, and the other pair is hole-like and curving downward, the $\varepsilon^{v}_\bk$ solution. We refer to these bands as the conduction band ($c$) and valence band ($v$), respectively. Furthermore, these conditions imply that the valence band states are $\ket{\frac32, m_j=\pm \frac32} $ angular momentum states and the conduction band states are $\ket{\frac32, m_j=\pm \frac12} $ states. (This may be seen by considering $h_\bk$ along $k_z$.) In this way, by tuning the chemical potential, we have access to a valence band Fermi surface consisting of pseudospin $\pm \frac32 $ states and a conduction band Fermi surface consisting of pseudospin $\pm \frac12 $ states. Due to the different axial angular momentum of states on these Fermi surfaces pairing is expected to affect them differently. Therefore, in our study of pairing gap structures we clearly distinguish between valence band and conduction band Fermi surfaces. In Sec. \ref{sec:qptopology}, where we present the detailed analysis of gap structures, we will focus on both these cases, with a special emphasis on the more intriguing case of a pseudospin $\pm \frac32$ Fermi surface.

The operators $f^\dagger_{\bk}$ create quasiparticles in valence band eigenstates, i.e., $f^\dagger_{\bk\mu} \ket{0} =  \ket{\bk, \mu; v}$, and similarly for $d^\dagger_{\bk}$. It would be desirable for these pseudospin operators to transform as ordinary spin under spatial and time-reversal symmetries. It is not guaranteed that such a basis for the band eigenstates exists.  We can, however, choose a basis such that the pseudospin states, $\ket{\bk, \up; v}$ and $\ket{\bk, \down; v}$ (in case of the valence band), transform as canonical Kramers partners under time-reversal and inversion symmetry. This justifies the (pseudo)spin labeling $\up,\down$, and implies that, when considering the pairing on the Fermi surface, we can speak of pseudospin-singlet and pseudospin-triplet pairing.

In the Luttinger Hamiltonian of Eq. \eqref{eq:hk} we have neglected the terms originating from the crystal field. As a result, $h_\bk$ is invariant under continuous joint spatial and spin rotations. This approximation may be justified at low energies close to the touching point when crystal anisotropy effects can be considered small. Furthermore, and perhaps more importantly, for our purpose of a gap structure classification of stationary pairing states it is natural to choose a starting point of higher symmetry. In fact, as will be shown in Secs. \ref{sec:GLtheory} and \ref{sec:qptopology}, a gap structure classification developed on the basis of a rotationally symmetric model naturally includes the analysis of pairing states with discrete crystal symmetry, since the symmetry group which leaves stationary points of the free energy invariant may in principle be any subgroup of the full rotation group. Therefore, our gap structure classification also applies to pairing states arising in cubic models (see Sec. \ref{ssec:cubicsymmetry}). This implies, for instance, that our study bears a direct connection to the Bi-based half-Heusler superconductors, in particular YPtBi. We do note, however, that in this work we consider inversion symmetric systems with a twofold degenerate Fermi surface.

We note in passing that in this paper we exclusively focus on superconductivity, and assume the presence of a Fermi surface due to hole or electron doping; other ordering instabilities, relevant at the touching point, have been addressed in Refs. \onlinecite{savary14,herbut14,boettcher17b,goswami17}.

\subsection{Pairing channels and their symmetry \label{sec:pairingclass}}

The first step towards an analysis of pairing states is the identification of distinct pairing channels. Given the symmetry group $G$ of the normal state material, the irreducible pairing channels are classified by the representations of $G$. In the present case, the Luttinger Hamiltonian of Eq. \eqref{eq:hk} has both time-reversal and inversion symmetry, and is invariant under joint rotations of spatial and spin degrees of freedom. As a result, including $U(1)$ charge conservation, the symmetry group can be written as $G=   U(1) \times SO(3) \times P \times  \Theta$. The irreducible pairing channels can be distinguished by the angular momentum quantum numbers of the Cooper pairs. The total angular momentum $J$ is the sum of the Cooper pair orbital angular momentum $L$ and spin angular momentum $S$; as a consequence of spin-orbit coupling the symmetry quantum numbers of the Cooper pairs are $(L,S;J,M_J)$, where $M_J$ is the magnetic quantum number describing the axial angular momentum. 

Pairing channels with nonzero $J$ have $2J+1$ independent components, transforming as partners under rotations, and are called multicomponent channels. The components are degenerate right at the superconducting transition temperature $T_c$: the transition temperature is a property of the channel and symmetry requires the symmetry-related components to have the same $T_c$. Our study of quasiparticle spectra and gap structures will require explicit expressions for these gap function components; they can be obtained using the standard $L,S$-coupling scheme for addition of angular momenta \cite{fang15,savary17}, as we will now briefly describe. 

In the case of $j=\frac32$ quasiparticles the total spin of the Cooper pair can take the values $S= 0,1,2,3$. Cooper pairs in a total spin $S=0$ and $S=1$ state are conventionally called singlet and triplet pairing states; by analogy $S= 2,3$ states can be called quintet and septet pairings. We define $\Pi^\dagger_{SM_S}(\bk)$ as the creation operator of a pair of quasiparticles with momenta $\bk$ and $-\bk$, in a state with total spin $S$ and magnetic quantum number $M_S$; $\Pi^\dagger_{SM_S}(\bk)$ is given by
\be
\Pi^\dagger_{SM_S}(\bk)  =   c^\dagger_{\bk\alpha} ( \mathsf{ S}_{SM_S} \Tm )_{\alpha\beta} c^\dagger_{-\bk\beta}. \label{eq:pairSMS}
\ee
Here, the matrices $ \mathsf{ S}_{SM}$ are the multipole matrices of spin $j =\frac{3}{2}$ fermions \cite{li16} and the anti-symmetric matrix $\Tm=e^{i\pi S_y}$ plays the role of $\epsilon\equiv is_y$ familiar from spin-$\frac{1}{2}$ pairing. 

The $S=0$ matrix is proportional to the identity, i.e., $ \mathsf{ S}_{00} =1/2$, and corresponds to a rotationally invariant spin-singlet pairing. The $S=1$ matrices transform as a magnetic dipole (i.e., pseudovector) and are given by linear combinations of the spin matrices $\bS$. The $S=2,3$ matrices are higher order multipole matrices, describing spin quadrupolar and octupolar pairing, respectively, and transform as rank-2 and rank-3 tensors. Together these matrices span the space of Hermitian $4\times 4$ matrices. (A more detailed discussion of the multipole matrices, including an explicit construction, can be found in Appendix \ref{app:spinmat}.) 

The internal spatial structure of the Cooper pair is captured by the orbital part of the Cooper pair wave function and is given by the spherical harmonics $Y_{LM_L}(\hat{\bk})$, where $\hat \bk = \bk/|\bk|$. In the familiar nomenclature, superconductors with orbital angular momentum $L=0,1,2,3$ are referred to as $s$-, $p$-, $d$-, and $f$-wave, respectively. The spin and orbital angular momenta of Cooper pairs are constrained by Fermi statistics: since the matrices $\mathsf{ S}_{SM_S}$ are symmetric (anti-symmetric) for even (odd) $S$, $S$ and $L$ must either both be even or both be odd. To form irreducible pairings, let $\Pi^\dagger_{JM_J}(\bk)$ be the operator which creates a pair of electrons in a state of total angular momentum $J$ and $M_J$. Such irreducible pair creation operators take the general form
\be
\Pi^\dagger_{JM_J}(\bk)  =    c^\dagger_{\bk\alpha} [ \mathsf{ J}_{JM_J}(\hat\bk) \Tm ]_{\alpha\beta} c^\dagger_{-\bk\beta}, \label{eq:pairJ}
\ee
where now the momentum-dependent matrices $ \mathsf{ J}_{JM_J}(\hat\bk)$ are linear combination of $Y_{LM_L}( \hat{\bk})$ and the spin matrices $\mathsf{ S}_{SM_S}$. The appropriate linear combinations are uniquely determined by the Clebsch-Gordan coefficients; specifically one has
\begin{multline}
 \mathsf{ J}_{JM_J}(\hat\bk)  = \\
\sum_{M_L+M_S=M_J} \bracket{LS;M_LM_S}{LS; JM_J} Y_{LM_L} (\hat\bk) \mathsf{ S}_{SM_S}, \label{eq:pairJMJ}
\end{multline}
where $\bracket{LS;M_LM_S }{LS;JM_J}$ are the Clebsch-Gordan coefficients. Then, we may then write down a pairing Hamiltonian $H^{(L,S;J)}_\Delta$ for pairing in the $(L,S;J)$ channel as
\be
H^{(L,S;J)}_\Delta  =  \sum_{\bk,M_J} \Delta_{M_J}\left(\frac{k}{k_F}\right)^L \Pi^\dagger_{JM_J}(\bk)  + \text{H.c.}. \, \label{eq:HpairJ}
\ee
The complex expansion coefficients $ \Delta_{M_J}$ define the multicomponent pairing order parameter; superconductors with total angular momentum $J$ are described by a $2J+1$-component order parameter. Different order parameter configurations with equal norm generally define different pairing states, with different symmetry properties (unless, clearly, the two configurations are related by a global rotation). A phenomenological theory of the superconducting order parameter will be developed in the next section, where ground state solutions of the free energy and their symmetry breaking patterns are discussed. 

\begin{table}[t]
\centering
\begin{ruledtabular}
\begin{tabular}{cll}
$J$  & \multicolumn{2}{l}{Combinations of $(L,S)$ such that $L+S=J$}\\ 
  & Even parity & Odd parity \\ [4pt]
\hline
$0$ & $(0,0)$, $(2,2)$ &  $(1,1)$, $(3,3)$  \\[5pt]
$1$ & $(2,2)$ &  $(1,1)$, $(3,3)$  \\[5pt]
$2$ & $(0,2)$,$(2,0)$, $(2,2)$ & $(1,1)$, $(1,3)$, $(3,3)$ \\[5pt]
$3$ & $(2,2)$ & $(1,3)$, $(3,3)$ \\[5pt] 
$4$ & $(2,2)$ & $(1,3)$, $(3,3)$  
\end{tabular}
\end{ruledtabular}
 \caption{{\bf Lowest order pairing multiplets.} Table summarizing the pairing multiplets $(L,S;J)$ up to $L=2$ and $J=4$. The leading and subleading order $s$- and $p$-wave pairings (i.e., $L=0$ and $L=1$) are most relevant when considering pairing instabilities of topological semimetals. The pairing functions $\mathsf{ J}_{JM_J}(\hat\bk) $ of the latter pairing channels, see Eq. \eqref{eq:pairJMJ}, can be found in Ref. \onlinecite{savary17}.
 }
\label{tab:classification}
\end{table}

In Table \ref{tab:classification} we list a number of $(L,S;J)$ pairing multiplets for orbital angular momenta up to $L=2$. The pairing functions $\mathsf{ J}_{JM_J}(\hat\bk) $ corresponding to the latter pairing channels up to $L=1$ are tabulated in Ref. \onlinecite{savary17}. For well-screened short-ranged interactions, the $s$-wave ($L=0$) and $p$-wave ($L=1$) pairing channels are expected to have highest $T_c$, with higher angular momentum $L$ channels being suppressed. The relative strength of $s$-wave and $p$-wave pairing instabilities is a question of high current interest, which we will not address here. For the purpose of this work we will take the position that both $s$-wave and $p$-wave pairing are likely to be relevant for the experimental systems under study.

It is useful to consider the symmetries of the pairing matrices $ \mathsf{ J}_{JM_J}(\hat\bk)$ in more detail. In particular, the transformation properties under $\Theta$ and $P$ will be of interest. First note that $\Theta$ and $P$ act on the quasiparticle-operators as 
\be
\Theta c_{\bk\alpha} \Theta^{-1}= \Tm_{\alpha\beta} c_{-\bk\beta}, \quad P c_{\bk\alpha} P^{-1}= c_{-\bk\alpha} ,\label{eq:symmetries}
\ee
where $\Tm= e^{i\pi S_y}$ is the time-reversal matrix of Eq. \eqref{eq:pairSMS}. (Time-reversal is defined as $\Theta = \Tm K$, where $K$ is complex conjugation.) This immediately implies that $P \spinS{S}{M_S} P^{-1} = \spinS{S}{M_S} $. Using that $Y_{LM_L}(-\hat\bk) = (-1)^L Y_{LM_L}(\hat\bk) $ one trivially obtains 
\be
P \mathsf{ J}_{JM_J}(\hat\bk) P^{-1} = (-1)^L \mathsf{ J}_{JM_J}(\hat\bk).  \label{eq:PsymJM}
\ee
The spin multipole matrices transform under time-reversal as $\Theta \spinS{S}{M_S} \Theta^{-1} = (-1)^{S+M_S}\spinS{S}{,-M_S} $, which can be derived from $\Tm {\bf S}^* \Tm^{\dagger}  = - {\bf S}$; for the spherical harmonics one has $\Theta Y_{LM_L}(\hat\bk) \Theta^{-1} = Y^*_{LM_L}(-\hat\bk) = (-1)^{L+M_L}Y_{L,-M_L}(\hat\bk)$. Combining this we find
\be
\Theta \mathsf{ J}_{JM_J}(\hat\bk) \Theta^{-1} = (-1)^{J+M_J} \mathsf{ J}_{J,-M_J}(\hat\bk).  \label{eq:TsymJM}
\ee
These symmetry properties, in combination with the transformation properties under $SO(3)$ rotations, are at the heart of both the phenomenological theory of multicomponent pairing and the subsequent gap structure analysis. We note that the pairings $\mathsf{ J}_{JM_J}(\hat\bk)$ are eigenstates of rotations about the $z$-axis. If $U_{{\theta z}}$ is the $j=\frac32$ spinor representation of a rotation $C_{\theta z} $ about the $z$-axis, then one has $U_{{\theta z}}\mathsf{ J}_{JM_J}( C^{-1}_{\theta z} \hat\bk) U^\dagger_{{\theta z}} = e^{-i\theta M_J}\mathsf{ J}_{JM_J}( \hat\bk)$, or, equivalently, $U^\dagger_{{\theta z}}\mathsf{ J}_{JM_J}( C_{\theta z} \hat\bk) U_{{\theta z}} = e^{i\theta M_J}\mathsf{ J}_{JM_J}( \hat\bk)$.

We conclude this section with two remarks. The first concerns the symmetry of the normal state. When cubic crystal anisotropy effects are important, the irreducible pairing channels are labeled by representations of the cubic point group. The number of distinct (i.e., orthogonal) pairing channels in systems with discrete crystal symmetry is finite. Since the dimension of cubic representations is at most three, the pairings of Eq. \eqref{eq:pairJ} are generally split into pairings with distinct cubic symmetry. For instance, assuming an inversion symmetric normal state, the $J=2$ channel is split into $J=2 \to  E_{g,u}+T_{2g,u}$, where $g$/$u$ denotes the parity of the pairing channel, i.e., even/odd under inversion; the pairing channels $J=3,4$ of Table \ref{tab:classification} are split as $J=3 \to A_{2g,u}+T_{1g,u}+T_{2g,u}$ and $J=4 \to A_{1g,u}+E_{g,u}+T_{1g,u}+T_{2g,u}$. The more precise splitting of the pairing components of Eq.~\eqref{eq:pairJMJ} into pairing functions transforming as partners of cubic representations is tabulated in Table \ref{tab:cubicsymmetry} of Appendix \ref{app:cubic}. For pairing channels up to $L=1$ such splitting has been worked out in \onlinecite{savary17}.

Furthermore, since the number of cubic representations is finite, pairings from distinct channels $(L,S;J)$ will collapse onto the same cubic channel. For instance, as may be seen from the splitting of the $J=2,3,4$ channels, the gap functions of cubic $T_{2g,u}$ pairing can have contributions from all three isotropic channels. As a result, gap functions of cubic pairings will be linear combinations of symmetry-allowed terms with coefficients not determined by symmetry, and in general can be quite complicated. Crucially, however, the symmetry group of a given pairing state, i.e., the subgroup of the normal state symmetry group which leaves a pairing state invariant, is manifest and independent of material-specific details. The symmetry group can be used to establish universal properties of gap functions which are independent of their specific form.

Second, we note that in an analysis of pairing one may choose to focus exclusively on the Fermi surface band (either valence or conduction band), and project out the ``high-energy'' band. Since the projected Fermi surface pseudspin operators only admit pseudospin-singlet and pseudospin-triplet pairing, $S=0$ and $S=2$ states cannot be distinguished on the Fermi surface, affecting the classification of irreducible pairings \cite{savary17}. For instance, $(L,S)=(0,2)$ pairing effectively collapses onto $(L,S)=(2,0)$ pairing on the Fermi surface \cite{brydon16,savary17}. In this work, in order to correctly capture the essential features of pairing gap structures, it is important to include the effect of pairing-induced coupling of conduction and valence band.

\section{Phenomenology of multicomponent pairing \label{sec:GLtheory}}

The Hamiltonian of Eq. \eqref{eq:HpairJ} describes Cooper pairing within an irreducible pairing channel $(L,S;J)$.  A particular pairing state is specified by the superconducting order parameter $\Delta_M$, which carries all information on its symmetry properties. (We drop the subscript $J$, as $M=M_J$ will always correspond to $J$ in what follows.) For multicomponent superconductors, the order parameter not only has an overall amplitude and phase, but also internal structure: different order parameter configurations generally correspond to distinct pairing states. These pairing states can be sharply distinguished by symmetry; multicomponent superconductors, in addition to $U(1)$ charge conservation, break symmetries such as time-reversal symmetry or rotation symmetry. 

The spontaneous breaking of symmetry due to the selection of a specific pairing state occurs below $T_c$. (Right at $T_c$ all states in a channel are degenerate). Consider, for instance, the simple example of a two-component $p$-wave order parameter $(p_x,p_y)$ in two dimensions. Below $T_c$ the system will select either one of two states as its ground state: a time-reversal even but anisotropic state, given by $\cos \theta p_x + \sin \theta p_y$ (where $\theta$ parametrizes a family of states), or a chiral pairing state of the form $p_x \pm i p_y $.

To find the order parameter configuration of the superconducting ground state, one minimizes the free energy of the superconductor, denoted $F_J$. At temperatures below but close to $T_c$, a phenomenological Ginzburg-Landau theory (GL) is applicable, and one may expand $F_J$ in powers of the order parameter and its gradients. The phenomenological GL expansion parameters determine the superconducting state below $T_c$. (At zero temperature an expansion is no longer valid, and the minimization must rely on the full free energy.) In general, for multicomponent orders, analytical minimization of the energy functional can become challenging or even untractable, as the number of interaction parameters increases with the number of components. It is not clear that a full analytical solution can be found when the number of components becomes large.  

A powerful and elegant alternative strategy to obtain the free energy minima is based on the observation that solutions corresponding to free energy minima typically have residual symmetry, i.e., they do not fully break the symmetry of the normal state. This has motivated the expectation that states with residual symmetry are primary ground state candidates. In fact, by deriving all states invariant under a subgroup of the full symmetry group (which may be continuous or discrete subgroups), it was demonstrated that it is possible to systematically identify stationary states of the energy functional \cite{michel80,bruder86}, which may then simply be compared by directly computing the energy. Even though there is no general proof that this delivers all stationary states, in all known cases where an analytical solution is available, the result matches the energy comparison of stationary states \cite{mermin74,bruder86,volovik85, ozaki85}. 

Two kinds of stationary states with residual symmetry exist. Inert states are stationary points of the free energy independent of the precise form of $F_J$ \cite{barton74,barton75, makela07, yip07, kawaguchi11}, whereas noninert states depend on the interaction parameters of the energy functional, requiring knowledge of its precise form. Following the method presented in Ref. \onlinecite{kawaguchi11} in the context of spinor Bose-Einstein condensates, both types of stationary states can be obtained from a symmetry classification of order parameter configurations. 

For the purpose of studying gap structures of multicomponent pairing states, which is the focus of this work, these considerations lead to the important conclusion that we can restrict to studying the class of stationary states. This is a significant simplification, since in practice (i.e., for paring channels with nonzero but small $J$), the set of stationary states is rather tractable. Furthermore, insofar as inert states are concerned, details of the energy functional are unimportant. In this section our aim is to describe how the stationary solutions of the free energy can be derived using symmetry principles, and address their symmetry properties. This will provide the foundation for our gap structure analysis in Sec. \ref{sec:qptopology}. We pay particular attention to the cases $J=1,2,3,4$. In this section, we also develop a Ginzburg-Landau theory for multicomponent pairing. This allows us to energetically compare the stationary solutions, and study how degeneracy lifting and spontaneous symmetry breaking can occur immediately below $T_c$.

\subsection{Symmetry properties and stationary pairing states \label{ssec:symmetries}}

The symmetry of the superconducting order parameter $ \Delta_M$ is determined by the pairings defined in Eq. \eqref{eq:pairJMJ}. For instance, it follows from Eqs. \eqref{eq:PsymJM} and \eqref{eq:TsymJM} that the order parameter transforms under $\Theta$ and $P$ as 
\beq
\Theta \; & : &  \; \Delta_M \to  (-1)^{M+J}\Delta^*_{-M}, \nonumber \\
P \; & : &\;   \Delta_M \to  (-1)^{L}\Delta_{M} .
\eeq
The transformation properties under $SO(3)$ rotations are uniquely fixed by the total angular momentum $J$. 

The symmetry of the order parameter can be made more transparent by adopting a representation for the pairing states which exploits the analogy with angular momentum states. Specifically, given the order parameter $\Delta_M$, we may write the pairing state $\ket{\bdelta}$ as
\be
\ket{\bdelta} = \sum_M \Delta_M \ket{J,M}. \label{eq:spinstate}
\ee
In this state vector representation, the states $\ket{J,M}$ are identified with the pairings $\spinJ{J}{M}$ of Eq. \eqref{eq:pairJMJ}. From this, it is then clear that the superconducting state transforms under rotations in a canonical way. In particular, if $R\in SO(3)$ is a rotation by an angle $\theta$ about an an axis $\bn$, then the rotated state $\ket{R\bdelta}$ is given by $\sum_{M'}(D_R)_{MM'}\Delta_{M'} $, where $D_R=\exp(-i\theta\bn\cdot \makebf{\mathcal{I}})$ is the matrix representation of $R$ and $\makebf{\mathcal{I}}=  (\mathcal{I}_x,\mathcal{I}_y,\mathcal{I}_z)^\trans$ are the angular momentum-$J$ generators of $SO(3)$. 

The state vector representation $\ket{\bdelta} $ also reinforces the interpretation of $\Delta_M$ as the Cooper pair wave function.  Collecting the components $\Delta_M$ in a $2J+1$-component vector, we can write $ \makebf{\Delta} = (\Delta_J,\ldots,\Delta_{-J})^\trans$. (Note, however, that for $J>1$ the order parameter transforms as a rank-$J$ tensor.) As is clear from Eq. \eqref{eq:spinstate}, a special class of pairing states arises when the Cooper pair wave function only has a single nonzero component, e.g., $\makebf{\Delta} = (0,\ldots,1,\ldots,0)^\trans$. The pairing states $\ket{J,M=0}$ are non-magnetic and time-reversal invariant up to an unimportant global $U(1)$ phase. For even (and nonzero) $J$ the $M=0$ states are nematic, whereas for odd $J$ these states are polar; the nematic and polar axes coincide with the quantization axis (i.e., the $z$-axis in this case). Both the nematic and polar states are invariant under rotations about the quantization axis. The pairing states $\ket{J,M\neq0}$, which have nonzero axial angular momentum, break time-reversal symmetry: their time-reversed partners are $\ket{J,-M}$. We will refer to these states as magnetic or \emph{chiral} pairing states. The chiral pairing states are eigenstates of rotations about the quantization axis with eigenvalue $e^{-i\theta M}$.

All pairing states given by $\ket{J,M}$ have the special property that they are inert states of the free energy \cite{makela07, yip07, kawaguchi11}: they are stationary points of the energy independent of its precise form. The states $\ket{J,M}$ have a continuous isotropy group, where the isotropy group is defined as the subgroup of total symmetry group $G$ which leaves the state invariant. In the case of the states $\ket{J,M}$ the isotropy group is isomorphic to $SO(2)$, i.e., the group of rotations about the quantization axis. Members of $SO(3)$ not part of the isotropy group generate different but symmetry-equivalent and energetically degenerate states. We may therefore take the $z$-axis as the axis of continuous rotations, without loss of generality. 

Additional inert states can be obtained by considering discrete isotropy subgroups of $SO(3)$; the discrete subgroups of $SO(3)$ are $C_n$ (cyclic group rotations by an angle $2\pi/n$ about a special axis), $D_n$ (dihedral group of $C_n$ and an additional orthogonal two-fold axis), $O$ (point group of the octahedron), $T$ (point group of the tetrahedron), $Y$ (point group of the icosahedron). An example of such a state is given by $\ket{\bdelta_{D_4}}= \ket{2,2}+\ket{2,-2}$, which is a pairing state of a $J=2$ superconductor with $D_4$ symmetry. 

In the process of constructing pairing states with discrete symmetry of a specific pairing channel one may find that some states are not uniquely determined. To obtain a stationary state of the free energy one minimizes the free energy over the manifold parametrizing these states. Such a stationary state is a noninert state, as it depends on the form of the energy.

Before we proceed to the GL theory, it is worth pointing out that instead of the magnetic basis used in Eq. \eqref{eq:spinstate}, one may write the pairing state $\ket{\bdelta}$ in a ``real'' basis as
\be
\ket{\bdelta} = \sum_{a} \Delta_a \ket{J,a}.  \label{eq:scstate}
\ee
Here, $\ket{J,a}$ are chosen such that time-reversal simply acts as $\ket{J,a} \to \ket{J,a}$, which implies for the order parameter $\Theta \; : \; \Delta_a \to \Delta^*_a$. For instance, for $J=1$ one has $a \in \{x,y,z\}$ and hence $ \bdelta=(\Delta_x,\Delta_y,\Delta_z)^\trans $; for $J=2$ one has $a \in \{x^2-y^2,3z^2-r^2,xz,yz,xy\}$. In this basis, time-reversal symmetry breaking pairing states are defined by order parameters which are not equal to their complex conjugates (up to a phase), which connects to standard treatments of multicomponent superconductivity. Note also that the rank-$J$ tensor structure of $\bdelta$ is particularly transparent in this basis.

\subsection{Ginzburg-Landau theory for general $J$\label{ssec:GLgeneral}}

For general angular momentum $J$, the free energy $F_J$ of the superconductor is an integral over the free energy density $f_J$,
\be
F_J = \int d^3\br \;f_{J}. \label{eq:FE}
\ee
In the GL regime (i.e., in the vicinity of $T_c$), where the strength of pairing is small, the free energy density $f_J$ can be expanded in powers of the order parameter $\bdelta$ and its gradients. At given order, the expansion consists of all independent terms invariant under the symmetries of the normal state. For our purposes it is sufficient to consider the homogeneous part of the free energy density and ignore contributions from spatial inhomogeneities. Up to fourth order in $\bdelta$, the free energy density can be expressed in the following general form
\be
 f_{J} = r  |\makebf{\Delta}|^2 +u |\makebf{\Delta}|^4+ \sum_K  v_{K} \sum^{K}_{N=-K}|I_{K N} |^2, \label{eq:FJ}
\ee
where $|\makebf{\Delta}|^2=\makebf{\Delta}^\dagger \makebf{\Delta}$. Here $r\propto (T-T_c)$, as is usual in GL theory. (In weak-coupling one has $r= n(\varepsilon_F)(T/T_c-1)$, where $n(\varepsilon_F)$ is the density of states at the Fermi energy.) Clearly, the first two terms only depend on the overall magnitude of $\bdelta$, and do not depend on the internal structure of the order parameter. ($|\makebf{\Delta}|^2$ is the only symmetry-allowed term at second order.) The third term is a sum over the magnitudes of the subsidiary order parameters $I_{K N}$, where $K$ is an angular momentum and $N$ the axial angular momentum. Subsidiary order parameters are bilinears (i.e., composites) of the superconducting order parameter $\bdelta$ and capture the broken symmetries of the superconducting state. In the present case, the subsidiary orders $I_{K N}$ describe the magnetic multipole moments of the superconductor. For instance, superconductors with nonzero $I_{1, N=1,0,-1}$ have a magnetic dipole moment and thus have a chirality; superconductors with nonzero $I_{2,N=2,\ldots,-2}$ have a magnetic quadrupole moment. It follows that the subsidiary order parameters encode the symmetry properties of the superconductor, and are thus sensitive to the internal structure of $\bdelta$. As a result, the GL coefficients $v_K$ are responsible for energetically discriminating different pairing states below $T_c$, and they favor (or disfavor) pairing states with a certain structure of multipole moments. 

As a result, the GL analysis of multicomponent superconductors is an analysis of the terms with interaction parameters $v_K$. We can write the subsidiary orders $I_{K N}$ as 
\be
I_{K N}  =  \makebf{\Delta}^\dagger   \mathcal{I}_{KN}  \makebf{\Delta} =\sum_{MM'}(\mathcal{I}_{KN})_{MM'}\Delta^*_M \Delta_{M'}, \label{eq:subsorder}
\ee
where $\mathcal{I}_{KN}$ are the corresponding multipole matrices of an angular momentum $J$, whose dimension $(2J+1)\times (2J+1)$ therefore depends on $J$. The structure of the matrices $\mathcal{I}_{KN}$ is discussed in more detail in Appendix \ref{app:multipole}. (Note that the $I_{K N} $ are gauge invariant, as is required for magnetic multipole order parameters.) As is clear from Eq. \eqref{eq:subsorder}, in the case of superconductors with Cooper pair angular momentum $J$, one can form $2J$ distinct subsidiary order parameters. Importantly, however, the sum over $K$ in Eq. \eqref{eq:FJ} can be restricted to $K=1,\ldots,J$. A proof of this, which builds on Ref. \onlinecite{barton74}, is provided in Appendix \ref{app:GLinvariant}. (Note that $K=0$ can be excluded as it would simply give another term $\propto |\bdelta|^4$.)

Since the subsidiary order parameters fully encode the symmetry properties of multicomponent pairing states, they can be used to uniquely distinguish classes of pairing states. More precisely, if two order parameter configurations represent the same pairing state, they are characterized by the same pattern of subsidiary orders, up to a global rotation. These states have the same structure of multipole moments. If, on the other hand, two stationary states of the free energy have distinct isotropy groups, and are therefore different pairing states, this will be reflected in their multipole moment signature; they will be associated with a different structure of subsidiary order. 

To establish a more concrete connection between the stationary states and subsidiary orders, consider, for instance, the pairing states $\ket{J,M}$. The chiral pairing states $\ket{J,M\neq 0}$ have a nonzero magnetic dipole moment proportional to $M$ along the $z$-axis. To see this, let us define 
\be
\bI = \begin{pmatrix}  I_x \\ I_y \\ I_z \end{pmatrix}  \label{eq:chirality}
\ee 
as the magnetic dipole moments along the $x,y,z$ axes, which is often referred to as chirality. These are related to $\{ I_{11},I_{10},I_{1-1} \}$ as $I_{z}= I_{10}$ and $I_{1\pm 1}= \mp(I_{x}\pm iI_{y})/\sqrt{2}$. Then, the chiral pairing states $\ket{J,M\neq 0}$ are characterized by $I_z \propto M$. 

Clearly, the pairing states $\ket{J,M= 0}$ have vanishing chirality. In fact, all time-reversal invariant pairing states must have vanishing magnetic multipole moments with odd $K$. Pairing states which do break time-reversal symmetry but have vanishing chirality belong to an exotic class of states with nonzero higher order odd-K multipole moment. The rotational symmetry breaking of the pairing states $\ket{J,M= 0}$ is reflected in a nonzero quadrupole moment (and, in general, higher order even-$K$ multipole moments). The quadrupole moment must be invariant under the rotations about the polar or nematic axis. 

We conclude the general discussion of GL theory with a remark regarding remaining degeneracies of the functional \eqref{eq:FJ}. In Eq. \eqref{eq:FJ} free energy density is expanded up to fourth order. While this is often sufficient to determine the ground state below $T_c$, it is possible that degeneracies (i.e., degeneracies of states with distinct symmetry) remain at this order, which only get lifted at sixth or higher order. If two symmetry distinct states are found to have the same energy at fourth order, one must take the GL expansion to the next order to find the ground state. Within the phenomenological GL theory, higher order terms (which are not simply products of lower order terms) are systematically and straightforwardly constructed using the multipole moment subsidiary order parameters. For instance, sixth order invariants are simply obtained by considering products $I_{K_1N_1}I_{K_2N_2}I_{K_3N_3}$ with $N_1+N_2+N_3=0$ and summing with the appropriate Clebsch-Gordan coefficients to generate total singlets.

\subsection{Examples: Application to $J=1,2,3,4$ \label{ssec:GLexamples}}

Let us consider some examples, starting with the simplest case $J=1$. Writing the order parameter in the real basis of Eq. \eqref{eq:scstate} as $\makebf{\Delta} =( \Delta_x , \Delta_y , \Delta_z  )^\trans$, the free energy density $f_{J=1}$ is given by
\be
 f_{J=1} = r  |\makebf{\Delta}|^2 +u |\makebf{\Delta}|^4+ v_{1}\sum_{M}|\makebf{\Delta}^\dagger \mathcal{I}_{1M }  \makebf{\Delta}|^2.  \label{eq:FJ1}
\ee
The superconducting state below $T_c$ is controlled by a single GL interaction coefficient $v_1$, giving rise to two possible ground states \cite{ho98}. The first is a non-magnetic polar state favored when $v_1>0$; the second, favored when $v_1<0$, is chiral and maximizes the magnetic dipole moment. 

To see this more clearly, consider the chirality $\bI$ of Eq. \eqref{eq:chirality}, where $I_a= \makebf{\Delta}^\dagger \mathcal{I}_a  \makebf{\Delta}$ with $\mathcal{I}_{z}= \mathcal{I}_{10}$ and $\mathcal{I}_{1\pm 1}= \mp(\mathcal{I}_{x}\pm i\mathcal{I}_{y})/\sqrt{2}$. The matrix elements of $\mathcal{I}_{x,y,z}$ have a very simple form, given by
\be
(\mathcal{I}_{a})_{bc} = -i \epsilon_{abc},  \qquad a,b,c \in \{x,y,z\}.   \label{eq:Ijkl}
\ee
As a result, one finds $\bI = (I_x,I_y,I_z)^\trans = -i\makebf{\Delta}^* \times \makebf{\Delta}$. In units where the pairing amplitude is set to 1, i.e., $|\bdelta|^2=1$, the solutions of Eq \eqref{eq:FJ1} are given by $\bI=0$ ($v_1>0$) and $|\bI|=1$ ($v_1<0$). In terms of Eq. \eqref{eq:spinstate} these solutions are simply expressed as the polar pairing state $\ket{1,0}$ and the chiral pairing state $\ket{1,1}$. (Note that $\ket{1,-1}$ is related to $\ket{1,1}$ by a twofold rotation about the $x$-axis.)

Next, we turn to the five-component $J=2$ superconductor. The corresponding five-component order parameter is defined as
\be
\makebf{\Delta} =( \Delta_{x^2-y^2} , \Delta_{3z^2-r^2} , \Delta_{yz} , \Delta_{zx} , \Delta_{xy}  )^\trans,  \label{eq:OPJ2}
\ee
and the free energy density $f_{J=2}$ takes the form
\begin{multline}
f_{J=2} =  r |\makebf{\Delta}|^2 +u |\makebf{\Delta}|^4 + v_1\sum_{M}|\makebf{\Delta}^\dagger  \mathcal{I}_{1M}  \makebf{\Delta}|^2\\
+ v_2\sum_{M}|\makebf{\Delta}^\dagger  \mathcal{I}_{2M}  \makebf{\Delta}|^2 ,
 \label{eq:FJ2}
\end{multline}
where now $v_{1,2}$ are two GL interaction coefficients. This GL energy functional was analytically solved by Mermin \cite{mermin74}, who studied a purely orbital $L=2$ $d$-wave order parameter, and was reconsidered by Sauls and Serene in the context of an $(L,S;J)=(1,1;2)$ (or $^3P_2$) order parameter for massive neutron stars \cite{sauls78}. 

To consider the free energy minima, it is convenient for the present purpose to follow the symmetry classification approach of Ref. \onlinecite{kawaguchi11}. Of the stationary states of $f_{J=2}$ only a subset of four corresponds to minima; the remaining states are saddle points. Two of these minima are inert states with a continuous symmetry group $SO(2)$. They are given by $\ket{\makebf{\Delta}_2} = \ket{2,2}$ and $\ket{\makebf{\Delta}_0} = \ket{2,0}$, or alternatively, in the real basis of Eq.  \eqref{eq:OPJ2} by
\be
\makebf{\Delta}_2 =(1 , 0,0,0,i )^\trans /\sqrt{2}.  \label{eq:J2state2}
\ee
and
\be
\makebf{\Delta}_0 =(0 , 1,0,0,0 )^\trans /\sqrt{2}.  \label{eq:J2state0}
\ee
Whereas the former is chiral and has $\bI = (0,0,I_z)^\trans= (0,0,2)^\trans$, the latter state is nematic with vanishing dipole moment. 

The two remaining states corresponding to free energy minima have discrete symmetry: the state $\ket{\makebf{\Delta}_T} = (\ket{2,2}+\ket{2,-2})/2+i\ket{2,0}/\sqrt{2}$ has tetrahedral symmetry and the state $\ket{\makebf{\Delta}_{D_4}} = (\ket{2,2}+\ket{2,-2})/\sqrt{2}$ has dihedral $D_4$ symmetry. In the real basis these take the form
\be
\makebf{\Delta}_T =(1 , \sqrt{2}i,0,0,1 )^\trans/2,  \label{eq:J2stateT}
\ee
and
\be
\makebf{\Delta}_{D_4} =(1 , 0,0,0,0 )^\trans, \label{eq:J2stateD}
\ee
respectively. Neither of these two states has a magnetic dipole moment: $\bI=0$ for both states. The tetrahedral state nevertheless breaks time-reversal symmetry, as is signaled by the relative phase in Eq. \eqref{eq:J2stateT}, and this is reflected in a nonzero octupole moment $I_{3N}$. The dihedral state is time-reversal invariant and its lowest nonzero multipole moment is quadrupolar. 

Even though they have different symmetry, the nematic state \eqref{eq:J2state0} and the dihedral state are similar in the sense that they both are time-reversal invariant and have nonvanishing quadrupole moment $I_{2N}$. In fact, these states are known as the uniaxial and biaxial nematic states, and they remain degenerate in energy to fourth order in the GL expansion \cite{mermin74}. The lifting of this degeneracy occurs at higher order in the expansion. That such a degeneracy lifting should occur is expected from the gap structures of these states, as we will demonstrate in the next section. Within weak-coupling BCS theory, it was found that the uniaxial nematic pairing is favored below $T_c$ \cite{boettcher17}.

Finally, we briefly discuss the cases $J=3$ and $J=4$. The free energy densities follow directly from Eq. \eqref{eq:FJ} and are straightforward generalizations of Eqs. \eqref{eq:FJ1} and \eqref{eq:FJ2}. Here, we will not quote their expressions explicitly, but instead focus our attention on the set of stationary states in these higher angular momentum channels. Clearly, the pairing states $ \ket{3,M}$ and $\ket{4,M}$ are stationary states with a continuous symmetry group. In addition to these, a number of stationary states with discrete symmetry and total angular momentum $J=3$ and $J=4$ exist. To illustrate this, let us consider the inert stationary states with discrete symmetry. For $J=3$ there are two such states, with octahedral $O$ and dihedral $D_6$ symmetry, given by
\beq
\ket{\bdelta_{O}} &=&  \ket{3,2} - \ket{3,-2},\label{eq:J3O} \\
\ket{\bdelta_{D_6}} &=& \ket{3,3} + \ket{3,-3} .\label{eq:J3D6}
\eeq
In the case of $J=4$ pairing, there are one octahedral, one tetrahedral, and three dihedral states, given by
\beq
\ket{\bdelta_{O}} &=&  \sqrt{5}\ket{4,4}  + \sqrt{14}\ket{4,0}+\sqrt{5} \ket{4,-4} ,\label{eq:J4O} \\
\ket{\bdelta_{T}} &=& \sqrt{7}\ket{4,4} +2i\sqrt{3} \ket{4,2}-\sqrt{10}\ket{4,0} \nonumber \\
&& +2i\sqrt{3}\ket{4,-2} + \sqrt{7} \ket{4,-4}, \label{eq:J4T}  \\
\ket{\bdelta_{D_8}} &=& \ket{4,4} - \ket{4,-4} ,\label{eq:J4D8}\\
\ket{\bdelta_{D_6}} &=& \ket{4,3} - \ket{4,-3} ,\label{eq:J4D6}\\
\ket{\bdelta_{D_4}} &=& \ket{4,2} + \ket{4,-2} .\label{eq:J4D4}
\eeq
(Note that we have not been concerned with normalization.) The tetrahedral state has the same symmetry as the state of Eq. \eqref{eq:J2stateT}. Except for this tetrahedral state, all these inert states correspond to time-reversal invariant pairing states. Apart from inert states, the $J=3,4$ pairing channels admit a number noninert states, which we will not list exhaustively here. An example will be considered in Sec. \ref{sec:qptopology}, where we present our gap structure analysis and classification; in particular, the gap structures of all inert states listed here will be considered.

Insofar as the energetics of these $J=3,4$ pairing states with discrete symmetry is concerned, we make a general observation. Since octahedral (i.e., cubic) symmetry forbids a quadrupole moment, and time-reversal symmetry forbids both a dipole and an octupole moment, the octahedral states $\ket{\bdelta_{O}} $ have vanishing $\sum_N|I_{KN}|^2$ for $K=1,2,3$ [see Eq. \eqref{eq:FJ}]. For the case $J=3$, this implies that the octahedral pairing state minimizes the GL free energy when $v_{1,2,3}>0$. Similarly, for the case $J=4$ it is possible to show that the octahedral pairing state minimizes the GL free energy when $v_{1,2,3}>0$ and $v_4<0$, since the octahedral state maximizes the total hexadecapole moment $\sum_N|I_{4N}|^2$. This observation can be viewed as an example of the general utility of expressing the energy functional in terms of quantities directly reflecting the pairing symmetry.

\section{Quasiparticle gap structures and topology \label{sec:qptopology}}

Based on the analysis of possible pairing ground states, in this section we turn to a detailed analysis of their quasiparticle gap structures, where we focus in particular on the associated topological quantum numbers. In three dimensions, four generic types of pairing gap structures can be distinguished: \emph{(i)} a full pairing gap; \emph{(ii)} bulk point nodes (codimension-3 nodes); \emph{(iii)} bulk line nodes (codimension-2 nodes); and \emph{(iv)} Bogoliubov Fermi surfaces (codimension-1 nodes). Gap structures of the latter kind have recently been introduced in Ref. \onlinecite{agterberg17}, where it was shown that these nodal degeneracies of codimension-1 are topologically stable in parity-even multiband superconductors with spontaneously broken time-reversal symmetry. 

Bulk point nodes correspond to Berry curvature monopoles in momentum space and must therefore come in pairs of opposite monopole charge \cite{nielssen83,haldane04}. In superconductors particle-hole symmetry ($\Xi$) imposes the constraint that a point nodal degeneracy at momentum $\bK$ on the Fermi surface must have a partner at $-\bK$ (e.g., the antipodal point on a spherical Fermi surface) with opposite monopole charge. If the quasiparticle spectrum consists of a single pair of point nodes, or more generally multiple \emph{non-degenerate} pairs, the low-energy gapless quasiparticles obey the Majorana equation of motion and realize itinerant Majorana fermions in three dimensions \cite{venderbos16,kozii16,meng12}. 

Different nodal gap structures arise when symmetries force pairs of point nodes to be degenerate. For instance, when time-reversal symmetry $\Theta$ is present each point node must be degenerate with a node of opposite Berry monopole charge. Such point nodes can be called Dirac points, by analogy with Dirac semimetals, realizing Dirac superconductors \cite{yang14}. A second kind of degenerate point nodes occurs when the degenerate nodes have the same monopole charge, as is the case in the canonical example of the superfluid $^3$He-A phase \cite{volovik}. 

These general considerations demonstrate that the topological properties of the quasiparticle spectrum are inextricably linked to the symmetry of superconducting state, as symmetries can put constraints on the gap structure. In particular, the parity of the superconducting state plays a crucial role: time-reversal invariant topological superconductors (with a full pairing gap) must have odd-parity pairing \cite{fu10,sato10}. Similarly, the parity of the pairing state is known to determine whether line nodes are stable degeneracies \cite{blount85,kobayashi14}. 

An analysis of gap structure topology must therefore clearly discriminate superconducting states with different symmetry. Accordingly, our derivation and classification of topological pairing states is built on the symmetry classification of stationary pairing states presented in the previous section. 

The organization of this section reflects this. We begin by both reviewing and establishing a number of general implications of symmetry-mandated constraints on gap structures. Armed with these, we then carefully examine the gap structures of: single-component $J=0$ superconductors (Sec. \ref{ssec:gapJ0}), multicomponent $M=0$ pairing states (Sec. \ref{ssec:gapnematic}), multicomponent chiral pairing states (Sec. \ref{ssec:gapchiral}), and, finally, pairing states with discrete symmetry (Sec. \ref{ssec:gapdiscrete}). 


To describe and study the quasiparticle gap structures of pairing states we adopt the mean-field formalism and define the Nambu spinor 
\be
\chi_\bk = \begin{pmatrix} c_{\bk} \\ \Tm c^{\dagger \trans}_{-\bk} \end{pmatrix}. \label{eq:nambu}
\ee
The superconducting mean-field Hamiltonian then takes the form
\be
\mathcal{H} = \frac{1}{2}\sum_\bk\chi^\dagger_\bk\mathcal{H}_\bk  \chi_\bk, \label{eq:HBdG}
\ee
with $\mathcal{H}_\bk$ given by
\be
\mathcal{H}_\bk =  \begin{pmatrix} h_\bk & \Delta_\bk \\  \Delta^\dagger_\bk & -h_\bk \end{pmatrix} = h_\bk\tau_z + \Delta_\bk\tau_++\Delta^\dagger_\bk\tau_- . \label{eq:HBdGk}
\ee
Here, $h_\bk$ is the Luttinger Hamiltonian of Eq. \eqref{eq:hk}, $\Delta_\bk$ is the pairing potential, and we have introduced a set of Pauli matrices $\tau_{z}$ and $\tau_{\pm}= (\tau_x \pm i \tau_y)/2$ acting on Nambu space. The pairing potential $\Delta_\bk$ follows from the pairing Hamiltonian in Eq. \eqref{eq:HpairJ} and is given by 
\be
\Delta_\bk  = \left(\frac{k}{k_F}\right)^L\sum_{M} \Delta_{M} \mathsf{J}_{JM}(\hat\bk) \label{eq:deltak}
\ee
As stated earlier, we will focus on the gap structures of order parameter configurations $\makebf{\Delta} = ( \Delta_{J}, \ldots, \Delta_{-J})^\trans$ corresponding to the possible mean-field ground states which were obtained in the previous section; pairing states that do not correspond to free energy extrema are not considered. 

At this stage, it is useful to consider the discrete symmetry properties of the Hamiltonian $\mathcal{H}_\bk$. $\mathcal{H}_\bk$ possesses a manifest particle-hole symmetry defined as $\Xi = \Cm K$, where $K$ is complex conjugation and $\Cm$ is a unitary matrix given by
\be
\Cm= \begin{pmatrix}  & \Tm^\dagger \\  \Tm &    \end{pmatrix}.
\ee
Specifically, the Hamiltonian satisfies
\be
\Cm \mathcal{H}^*_\bk \Cm^{\dagger} =-\mathcal{H}_{-\bk}. 
\ee
Furthermore, depending on the parity of the orbital angular momentum $L$, the pairing potential is either even or odd under inversion, i.e., 
\be
P \Delta_\bk P^{\dagger} = \pm\Delta_{-\bk},
\ee
where $P$ acts as the identity. For odd-parity pairing states, the inversion can be redefined as $\Pm$ acting as $\tau_z$, such that the Hamiltonian $\mathcal{H}_\bk$ is inversion-symmetric, $\Pm \mathcal{H}_\bk \Pm^{\dagger} =\mathcal{H}_{-\bk}$. This implies, however, that $\Pm$ and $\Xi$ do not commute, but instead satisfy the anticommutation relation $\{ \Xi, \Pm \} =0$. 

A time-reversal symmetric pairing potential satisfies $\Tm \Delta^*_\bk \Tm^{\dagger} = \Delta_{-\bk}$. Since the pairing potential also obeys Fermi statistics, expressed as $\Delta_\bk = \Delta^\trans_{-\bk}$, time-reversal invariance implies that the pairing potential is Hermitian: $\Delta^\dagger_\bk=\Delta_\bk$. As a result, time-reversal invariant superconductors in three dimensions (Altland-Zirnbauer class DIII) admit a $\mathbb{Z}$ topological classification in terms of a winding number \cite{schnyder08,qi10}. Any improper spatial symmetry, i.e., an inversion or mirror symmetry, forces the winding number to be zero, and this has lead to the important insight that time-reversal invariant topological superconductors in three dimensions must have odd-parity pairing symmetry \cite{fu10,sato10}. In particular, when the Fermi surface (or, more generally, the set of disconnected Fermi surfaces) enclose an odd number of time-reversal invariant momenta, a fully gapped odd-parity superconductor is a topological superconductor. This is a powerful corollary which we can directly apply to the present case where we consider a single (either valence or conduction band) Fermi surface around $\Gamma$.

The mean-field Hamiltonian of Eq. \eqref{eq:HBdGk} is expressed in the orbital basis; since we are interested in pairing on the Fermi surface it is advantageous rewrite it in the band basis, defined by the $f_{\bk}$ and $d_{\bk}$ operators. The quasiparticle operators $c_{\bk}$ and $c^\dagger_{\bk}$ can then be expressed in terms of $f_{\bk}$ and $d_{\bk}$ as
\be
c_{\bk}  =  V_\bk f_{\bk} + W_\bk d_{\bk}, \quad  c^\dagger_{\bk}  =  f^\dagger_{\bk} V^\dagger_\bk + d^\dagger_{\bk}W^\dagger_\bk ,
 \label{eq:ctofd}
\ee
where the matrices $V_\bk$ and $W_\bk$ contain the eigenvectors of the valence band and conduction band states, respectively. We choose a basis such that the Fermi surface pseudospin degrees of freedom transform under $\Theta$ and $P$ as $\Theta f_{\bk\mu} \Theta^{-1}= \epsilon_{\mu\nu} f_{-\bk\nu}$ and $P f_{\bk\mu} P^{-1}= f_{-\bk\mu}$. (See appendix \ref{app:projection} for explicit expressions.) Using Eq. \eqref{eq:ctofd} we rewrite $H$ in the band basis as 
\be
\mathcal{H} = \frac{1}{2}\sum_\bk \begin{pmatrix} \psi^\dagger_{\bk } & \varphi^\dagger_{\bk }   \end{pmatrix}  \begin{pmatrix} \mathcal{H}^{vv}_\bk &  \mathcal{H}^{vc}_\bk \\  \mathcal{H}^{cv}_\bk &  \mathcal{H}^{cc}_\bk   \end{pmatrix} \begin{pmatrix} \psi_{\bk } \\  \varphi_{\bk }   \end{pmatrix},
\ee
where $\psi_{\bk }$ and $\phi_{\bk }$ are Nambu spinors for the band operators, i.e., 
\be
 \psi_{\bk } = \begin{pmatrix} f_{\bk } \\ \epsilon f^{\dagger \trans}_{-\bk } \end{pmatrix}, \quad  \varphi_{\bk } = \begin{pmatrix}   d_{\bk } \\ \epsilon d^{\dagger \trans}_{-\bk }  \end{pmatrix}.  \label{eq:fdnambu}
\ee
The projection of $\mathcal{H}_\bk$ onto the valence band, denoted $\mathcal{H}^{vv}_\bk $, is given by
\be
\mathcal{H}^{vv}_\bk =  \begin{pmatrix}  \varepsilon^{v}_{\bk} &  V^\dagger_\bk \Delta_\bk V_\bk \\   V^\dagger_\bk \Delta^\dagger_\bk V_\bk & - \varepsilon^{v}_{\bk}  \end{pmatrix} ,\label{eq:hvv}
\ee
and $\mathcal{H}^{cc}_\bk $ is simply obtained from the substitutions $v \leftrightarrow c$ and $V_\bk \leftrightarrow W_\bk$. The off-diagonal blocks, which represent a pairing-induced coupling of the valence and conduction bands, take the form
\be
\mathcal{H}^{vc}_\bk = (\mathcal{H}^{vc}_\bk)^\dagger = \begin{pmatrix}  0 &  V^\dagger_\bk \Delta_\bk W_\bk \\   V^\dagger_\bk \Delta^\dagger_\bk W_\bk & 0 \end{pmatrix}  \label{eq:hvc}
\ee
To describe pairing within the valence one can simply project out the conduction band and take Eq. \eqref{eq:hvv}. This ignores the effects of coupling to the conduction band captured by Eq. \eqref{eq:hvc} and potentially misses qualitative features of the gap structure with topological origin \cite{agterberg17}. Let us therefore take more formal approach which can account for all constraints imposed by the symmetry of the system. The resolvent corresponding to $\mathcal{H}_\bk$ takes the form
\be
\mathcal{G}(\bk,\omega) = (\omega-\mathcal{H}_\bk)^{-1} =\begin{pmatrix} \mathcal{G}^{vv} & \mathcal{G}^{vc}\\  \mathcal{G}^{cv} &  \mathcal{G}^{cc}  \end{pmatrix} . \label{eq:resolvent}
\ee
Using the properties of the resolvent, its valence band block is obtained as
\be
\mathcal{G}^{vv}(\bk,\omega) = [\omega-\tilde{\mathcal{H}}^{vv}_{\bk}(\omega)]^{-1}, \label{eq:Gvv}
\ee
where the effective Hamiltonian $\tilde{\mathcal{H}}^{vv}_{\bk}(\omega) $ is given by
\be
\tilde{\mathcal{H}}^{vv}_{\bk}(\omega) = \mathcal{H}^{vv}_\bk + \mathcal{H}^{vc}_\bk(\omega-\mathcal{H}^{cc}  _\bk)^{-1}\mathcal{H}^{cv}_\bk.  \label{eq:Hvveff}
\ee
The poles of \eqref{eq:Gvv} still give the exact eigenenergies as long as the corresponding eigenstates have nonzero support on the valence band states. Since pairing is typically assumed to involve states on the Fermi surface, and one is interested in small energies $\omega$ compared to the Fermi energy, the effective Hamiltonian can be expanded to lowest order in $\omega/\varepsilon^c$, where $\varepsilon^c$ is the energy of the conduction band at momenta on the Fermi surface. Importantly, a number of properties of the effective valence band pairing Hamiltonian \eqref{eq:Hvveff} can be established by simply invoking symmetry arguments. 

First, note that symmetries of the normal state, by definition, do not mix conduction and valence band states, implying that their action is block-diagonal in the $(\psi_{\bk } , \varphi_{\bk } )^\trans$ basis. Second, note that Eq. \eqref{eq:Hvveff} together with the poles of $\mathcal{G}^{vv}(\bk,\omega) $, implicitly defines the full effective pairing potential $\tilde\Delta_\bk$ of the valence band, where $\tilde\Delta_\bk$ is a $2\times 2$ matrix in pseudospin space. The symmetry properties of $\Delta_\bk$ [Eq. \eqref{eq:deltak}] carry over to $\tilde\Delta_\bk$. In particular, given our choice of pseudospin basis, for even-/odd-parity pairing one has $\tilde\Delta_\bk = \pm \tilde\Delta_{-\bk}$, and time-reversal symmetric pairing implies $\epsilon \tilde\Delta^*_\bk \epsilon^\trans =  \tilde\Delta_{-\bk}$. In combination with Fermi statistics time-reversal symmetry leads to a Hermitian pairing potential $\tilde\Delta^\dagger_{\bk}=\tilde\Delta_{\bk}$. 

Then, in the case of odd-parity pairing, $\tilde\Delta_{\bk}$ may be expanded in pseudospin Pauli matrices $s_{x,y,z}$ as $\tilde\Delta_{\bk} = {\bf g}(\bk)\cdot \bs$, where ${\bf g}(\bk) = {\bf g}^*(\bk) =-{\bf g}(-\bk)$ is real. We would like to consider constraints imposed on ${\bf g}(\bk)$ by mirror symmetries. To this end we must establish how the valence band pseudospin states transform under mirror symmetry. Using the pseudospin basis presented in Appendix \ref{app:projection}, we find that the pseudospin matrix representations $O_{M_x}$ and $O_{M_y}$ of the mirror symmetries $M_{x}\, : \, x\to -x$ and $M_{y}\, :\, y\to -y$ take the simple form
\be
O_{M_x} = -is_x, \quad O_{M_y} = -is_y.  \label{eq:mirrorsymmetry}
\ee
This proves that a pseudospin basis exists such that the pseudspin transforms as an ordinary spin under mirror symmetry, and immediately implies that mirror symmetry imposes constraints on the gap function. Specifically, on the mirror plane ${\bf g}$ must be normal to the mirror plane. Moreover, since the equation $g_x(\bk)=0$ has a one-parameter family of solutions on a $yz$ mirror plane, the valence band gap structure has mirror symmetry protected point nodes where the solutions of $g_x(\bk)=0$ intersect the Fermi surface \cite{blount85}. These points nodes must be degenerate due to time-reversal symmetry. 

We have thus obtained the result that time-reversal invariant odd-parity pairing states with a mirror symmetry generically have degenerate point nodes (with opposite Berry monopole charge). The point nodes are protected by mirror symmetry \cite{blount85,zhang13,yang14}.

\subsection{Pairing states with a rotation axis \label{ssec:rotaxis}}

\begin{figure}
\includegraphics[width=\columnwidth]{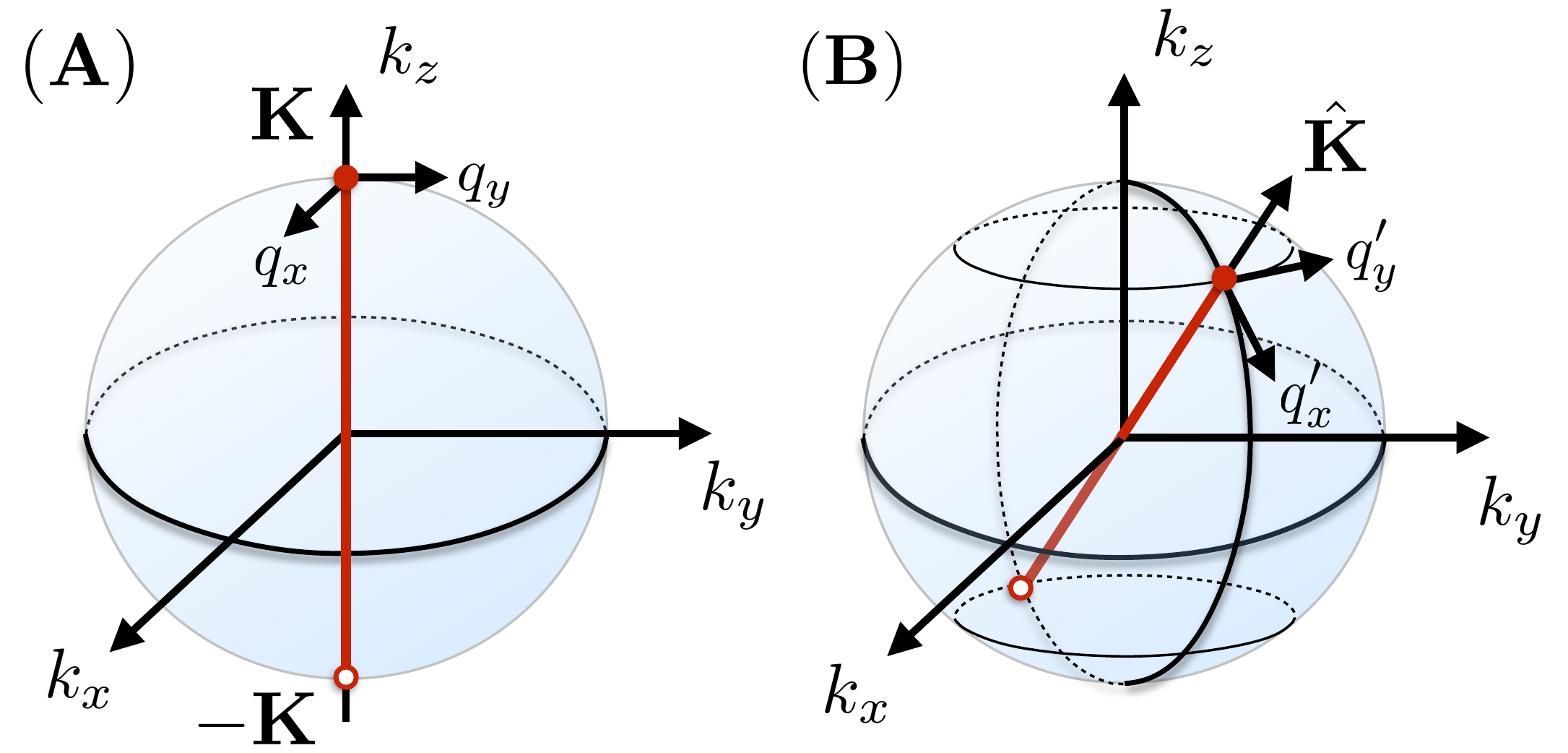}
\caption{\label{fig:figFS} {\bf Fermi surface and rotation axis.} ($\bf A$) Schematic representation of the Fermi surface (blue sphere) and the rotation axis along the $z$-direction (red solid line). In most cases, in particular in the case of the pairing states $\ket{J,M}$, one may take the rotation axis along $\hat z$. The Fermi surface momenta on the rotation axis are $\pm \bK = \pm k_{Fv} \hat z$ (or  $\pm k_{Fc} \hat z$ in case of the conduction band). ($\bf B$) In general, the axis of special rotational symmetry may be arbitrary. In the case of pairing states with discrete symmetry, there may be more than one discrete rotation axis, which is familiar from crystal point groups. 
 }
\end{figure}

We have learned from Sec. \ref{sec:GLtheory} that pairing states of multicomponent superconductors generically have special axes of rotation symmetry. These may be principal axes of continuous rotations or (a set of equivalent) discrete rotation axes. At Fermi surface momenta which lie on the rotation axis, rotation symmetry can give rise to constraints on the gap structure, leading to point nodal degeneracies and gapless quasiparticle excitations \cite{kozii16}. Therefore, to study the gap structure of superconductors with a rotation axis, we develop a symmetry-based theory for the quasiparticle spectrum in the vicinity of the rotation-invariant Fermi surface momenta, which we denote $\pm \bK$, see Fig. \ref{fig:figFS}. In most cases we will be able to choose $ \bK$ along the $z$-axis, without loss of generality, in which case $ \bK=  k_{Fv,c} \hat z$ (Fig. \ref{fig:figFS} A). Here, $\bK=  k_{Fv,c}$ is the Fermi momentum of a valence or conduction band Fermi surface, and is given by
\be
k_{Fc} = \sqrt{\frac{2m\mu}{\kappa_1+\kappa_2}}, \quad k_{Fv} = \sqrt{\frac{2m\mu}{\kappa_1-\kappa_2}}.
\ee

In the case of a valence band Fermi surface (a case we will often consider as an example), we expand the Nambu spinor $\psi_\bk$ in small momenta $\bq$ around $\pm \bK$ and define the spinor $\Psi_{\bq \mu } $ as
\be
\Psi_{\bq \mu } = \begin{pmatrix} f_{\bq 1 \mu} \\  f_{\bq 2  \mu} \\  f^\dagger_{-\bq 1  \mu}  \\  f^\dagger_{-\bq 2  \mu}  \end{pmatrix} \equiv \begin{pmatrix} f_{\bK +\bq \mu} \\  f_{-\bK +\bq   \mu} \\  f^\dagger_{\bK -\bq   \mu}  \\  f^\dagger_{-\bK -\bq   \mu}  \end{pmatrix} , \label{eq:Psiq}
\ee
where we have introduced the label $1,2$ for the nodal degree of freedom $\pm \bK$. Recall that for the valence band $\mu=\up,\down$ refers to the $\pm \frac32$ pseudospin states. Here, the quantization axis is chosen along the rotation axis defined by $\bK$, such that under rotations by an angle $\theta$ one has $C_{\theta}f_{\bK\up,\down}C^{-1}_{\theta} = e^{\pm i 3\theta/2}f_{\bK\up,\down}$. Similarly, in the case of a conduction band Fermi surface, we collect the conduction band degrees of freedom close to $\pm \bK$ in the spinor $\Phi_{\bq \mu}$, given by
\be
\Phi_{\bq \mu} =\begin{pmatrix} d_{\bq 1 \mu } \\  d_{\bq 2 \mu} \\  d^\dagger_{-\bq 1 \mu}  \\  d^\dagger_{-\bq 2 \mu}  \end{pmatrix} \equiv \begin{pmatrix} d_{\bK +\bq \mu} \\  d_{-\bK +\bq   \mu} \\  d^\dagger_{\bK -\bq   \mu}  \\  d^\dagger_{-\bK -\bq   \mu}  \end{pmatrix}.  \label{eq:Gammaq}
\ee
Note that now, however, the pseudospin label $\mu=\up,\down$ refers to the $\pm \frac12$ states, such that under rotations one has $C_{\theta}d_{\bK\up,\down}C^{-1}_{\theta} = e^{\pm i \theta/2}d_{\bK\up,\down}$.

Expanded in these degrees of freedom, the Luttinger Hamiltonian of Eq. \eqref{eq:hk} takes the form
\be
H_0 \simeq  \frac{1}{2}\sum_\bq  \Psi^\dagger_{\bq } h^{v,c}_\bq \Psi_{\bq }  + \frac{1}{2}\sum_\bq  \Phi^\dagger_\bq h^c_\bq \Phi_\bq,
\ee
depending on whether one is considering a valence band or conduction band Fermi surface. Here, $h^{v,c}_\bq$ are given by
\be
h^{v,c}_\bq  = \begin{pmatrix}  \varepsilon^{v,c}_{\bq}  &  &  &  \\  &  \varepsilon^{v,c}_{-\bq} & &  \\  & & -  \varepsilon^{v,c}_{-\bq} &   \\   && &  - \varepsilon^{v,c}_{\bq}\end{pmatrix} , \label{eq:Hqvc}
\ee
with $ \varepsilon^{v,c}_{\pm \bq} \equiv  \varepsilon^{v,c}_{\bK\pm \bq}$. Note that we have used the inversion symmetry of the normal state: $\varepsilon^{v,c}_\bk =\varepsilon^{v,c}_{-\bk}$.

Now, let us specifically consider a valence band Fermi surface. The dispersion $\varepsilon^{v}_{\pm  \bq}$ can be expanded in small $\bq$ as
\be
\varepsilon^{v}_{\pm  \bq} =\pm v_{Fv} \bq \cdot \hat\bK + \frac{\kappa_1-\kappa_2}{2m}( \hat\bK \times \bq)^2 ,
\ee
where $\hat\bK = \bK/k_{Fv}$ and $v_{Fv}$ is the valence band Fermi velocity. If $\pm \bK$ is along the $z$-axis, this reduces to 
\be
\varepsilon^{v}_{\pm  \bq} =\pm v_{Fv} q_z + \frac{\kappa_1-\kappa_2}{2m}(q_x^2+q_y^2).   \label{eq:Enormalstate}
\ee
Note that the conduction band constitutes a high-energy subspace, located at energy $\varepsilon^{c}_{\bK} = \mu(\kappa_1+\kappa_2)/(\kappa_1-\kappa_2)$.

Next, consider the pairing at and close to $\pm \bK$. We distinguish two cases: even-parity pairing and odd-parity pairing. Even-parity pairing states have total spin $S=0,2$ (see Sec. \ref{sec:pairingclass}), which, when projected onto the valence band, implies pseudospin-singlet pairing. To describe even-parity pairing in the vicinity of $\pm \bK$ we define the pseudospin-singlet operator $F^\dagger_{s}(\bq)$ as
\be
F^\dagger_{s}(\bq) = \frac{1}{\sqrt{2}}( f^\dagger_{\bq 1 \up}f^\dagger_{-\bq 2 \down} - f^\dagger_{\bq 1 \down}f^\dagger_{-\bq 2 \up} ). \label{eq:Fsinglet}
\ee
Odd-parity pairing states have total spin $S=1,3$, and consequently have pseudospin-triplet structure when projected onto the valence band. In accordance, we define the three pseudospin-triplet operators as 
\beq
F^\dagger_{t+}(\bq) &=&  f^\dagger_{\bq 1 \up}f^\dagger_{-\bq 2 \up},  \nonumber \\
F^\dagger_{t0}(\bq) &=&  \frac{1}{\sqrt{2}}( f^\dagger_{\bq 1 \up}f^\dagger_{-\bq 2 \down} +f^\dagger_{\bq 1 \down}f^\dagger_{-\bq 2 \up} ) ,\nonumber  \\
F^\dagger_{t-}(\bq) &=&  f^\dagger_{\bq 1 \down}f^\dagger_{-\bq 2 \down} .   \label{eq:Ftriplet}
\eeq
In addition to these pairing operators, in order to capture the full low-energy structure of pairing in the valence band, we define the following effective Zeeman-type spin-split operator
\be
F_{Z}(\bq) = \sum_{j=1,2} f^\dagger_{\bq j \up}f_{\bq j \up} -  f^\dagger_{\bq j \down}f_{\bq j \down}. \label{eq:Fzeeman}
\ee
Such effective splitting of the pseudospin states can arise as a result of a pairing-induced coupling of the conduction band and valence band, see Eq. \eqref{eq:hvc}, and therefore relies on the multiband nature of quadratic semimetal. Clearly, $F_{Z}(\bq)$ can only be present if time-reversal symmetry is broken. 

With the definition of these operators, the projected pairing Hamiltonian in the case of even-spin pairing takes the form
\be
H_\Delta \simeq \sum_{\bq} \left[  \Delta_{\bq s} F^\dagger_{s}(\bq)+ \text{H.c.}  \right] + \delta   \sum_{\bq}F_{Z}(\bq), \label{eq:LEpairinghamSeven}
\ee
whereas for odd-spin pairing it is given by
\begin{multline}
H_\Delta \simeq\sum_{\bq} \left[ \Delta_{\bq+} F^\dagger_{t+}(\bq) + \Delta_{\bq-} F^\dagger_{t-}(\bq) \right. \\
\left. + \Delta_{\bq 0} F^\dagger_{t0}(\bq) + \text{H.c.} \right] + \delta   \sum_{\bq} F_{Z}(\bq) . \label{eq:LEpairinghamSodd}
\end{multline}
It follows from \eqref{eq:Fzeeman} that $\delta \sim |\Delta|^2/\varepsilon^c_\bK$. The general program for the remaining parts of this section which pertain to pairings with a rotation axis is to derive constraints imposed by symmetry on effective low-energy gap functions. In appendix \ref{app:qpproject} we show how to obtain these gap functions from any particular pairing potential $\Delta_\bk$ [Eq. \eqref{eq:deltak}] by explicitly projecting onto the low-energy degrees of freedom.

\subsection{Spin-selective vs. spin-degenerate pairing \label{ssec:spinselective}}

In the familiar case of spin $j=\frac12$ pairing, Cooper pairs can be in a spin-singlet ($S=0$) or spin-triplet ($S=1$) state (assuming parity is a good quantum number). The quasiparticle spectrum of a spin-singlet superconductor is manifestly two-fold spin-degenerate. In contrast, spin-triplet superconductors are either unitary or non-unitary. Non-unitary superconductors necessarily break time-reversal symmetry and have the property that the two spin species have different quasiparticle spectra, i.e., the spectrum is not manifestly degenerate. The latter can have important implications for the gap structure, as it is a necessary condition for \emph{non-degenerate} point nodes to exist. We note in passing that the converse is not true: time-reversal symmetry breaking does not necessarily imply non-unitary pairing. In spin-orbit coupled systems, however, time-reversal symmetry breaking generically leads to non-unitary pairing, since symmetry-allowed terms in the gap function (which reflect spin-orbit coupling) generically give different pairing for spin-up and spin-down. 

The notion of degenerate and non-degenerate quasiparticle spectra can be generalized to pairing states of $j=\frac32$ fermions. We will call pairing states with a non-degenerate quasiparticle spectrum \emph{spin-selective} pairing states, and states with a manifestly two-fold degenerate spectrum \emph{spin-degenerate} pairing states. (Note that the distinction `unitary' versus `non-unitary' is specific to spin-$\frac12$ pairing.) Clearly, spin-selective versus spin-degenerate pairing pertains to the spin structure of the Cooper pairs. Spin-selective pairing states should be understood as states described by a pairing potential which contains $\spinS{S}{M_S}$ and $\spinS{S}{,-M_S}$ in an asymmetric way.  

Since time-reversal symmetry guarantees a two-fold degenerate spectrum, spin-selective pairing states must break time-reversal symmetry. In general, the converse is certainly not true. As in the aforementioned case of $j=\frac12$ systems, however, spin-orbit coupling generically leads to spin-selective pairing when time-reversal symmetry is broken. To see this in the present context, one may consider the irreducible spin-orbit coupled pairings given by Eq. \eqref{eq:pairJMJ}. A given pairing $\mathsf{J}_{JM}$ with $M\neq0$ is an asymmetric sum over orbital and spin angular momentum. For instance, the pairing $\mathsf{J}_{31} = c_1 Y_{11} \mathsf{S}_{30} +c_2 Y_{10} \mathsf{S}_{31} + c_3 Y_{1,-1} \mathsf{S}_{32}$ ($c_{1,2,3}$ are unimportant Clebsch-Gordan coefficients) contains $\mathsf{S}_{31}$ and $\mathsf{S}_{32}$ but neither $\mathsf{S}_{3,-1}$ nor $\mathsf{S}_{3,-2}$. This implies different pairing for spin species related by time-reversal symmetry and thus constitutes spin-selective pairing.

\subsection{Gap structures of single-component $J=0$ superconductors \label{ssec:gapJ0}}

For completeness we briefly review the total angular momentum $J=0$ pairings, which were considered and characterized in Refs. \onlinecite{fang15} and \onlinecite{li16}. Four different $J=0$ pairings exist, corresponding to combinations $(L,S)=(0,0),(1,1),(2,2),(3,3)$. All of these are time-reversal invariant and fully gapped. As a result, the odd-parity pairing states realize topological superconductors in class DIII and may be viewed as higher spin analogs $^3$He B-phase.  

They are, however, different from the $^3$He B-phase in a crucial way, which depends on the structure of the Fermi surface. Both Refs. \onlinecite{fang15} and \onlinecite{li16} have reported that in the case of a pseudospin $\pm \frac32$ Fermi surface (a valence band Fermi surface, in the present context) the $\mathbb{Z}$ winding number characterizing the odd-parity $J=0$ topological superconductor is $\pm 3$, making it topologically distinct form the $^3$He B-phase. This distinction is expressed through the surface state spectrum. 

The even-parity $J=0$ pairing states are topologically trivial; Table \ref{tab:J=0} summarizes the properties of $J=0$ pairings studied in Refs. \onlinecite{fang15} and \onlinecite{li16}. 

\begin{table}[t]
\centering
\begin{ruledtabular}
\begin{tabular}{ccc}
Parity & $(L,S)$ & Gap structure topology  \\ [4pt]
\hline  
Even & $(0,0)$, $(2,2)$ & Fully gapped, trivial in class DIII \\
Odd & $(1,1)$, $(3,3)$  & Fully gapped, class DIII topological 
\end{tabular}
\end{ruledtabular}
 \caption{{\bf Total angular momentum $J=0$ pairing.} Gap structure topology of fully gapped single-component pairing states with total angular momentum $J=0$. See Refs. \onlinecite{fang15} and \onlinecite{li16}.
 }
\label{tab:J=0}
\end{table}

\subsection{Gap structure of non-chiral $M=0$ superconductors \label{ssec:gapnematic}}

As discussed in Sec. \ref{ssec:symmetries}, the pairing states $\ket{J,0}$ are time-reversal invariant (up to a phase) and therefore spin-degenerate. As far as spatial symmetries are concerned, the states $\ket{J,0}$ can be distinguished by two symmetry quantum numbers: the parity of $L$ (i.e., even/odd under inversion) and the parity of $J$. The parity of $J$ determines whether $\ket{J,0}$ is even or odd with respect to twofold rotation about an axis perpendicular to the quantization axis. Without loss of generality we take the latter to be the $z$-axis. One then has that $\ket{J,0}$ is even (odd) under a twofold rotation about an axis in the $xy$-plane when $J$ is even (odd). This is why we may call even-$J$ states nematic and odd-$J$ states polar. 

Importantly, since mirror reflections in a plane that contains the $z$-axis are the product of inversion and perpendicular twofold rotations, the parity of $L$ and $J$ also determine the mirror symmetry properties of $\ket{J,0}$. The mirror symmetry properties have important consequences for the gap structures of both even-parity and odd-parity pairing states. 

We first consider odd-parity pairing (i.e., $L$ odd). Then, the pairing states $\ket{J,0}$ are even (odd) under mirror reflections in planes perpendicular to the $xy$-plane when $J$ is odd (even). For instance, the state $\ket{2,0}$ is odd under mirror symmetry. This directly leads us to an important observation: odd-parity pairing states $\ket{J,0}$ with even $J$ have neither an inversion symmetry nor a mirror symmetry, and, as a result, there are no constraints on the gap function which might enforce nodal degeneracies. We conclude that these pairing states have a full pairing gap on the Fermi surface and are time-reversal invariant topological pairing states in class DIII. Notably, these topological superconductors are different from the $J=0$ superconductors mentioned in Sec. \ref{ssec:gapJ0}, since the former break rotation symmetry and have a nematic axis. 

Due to the presence of a mirror symmetry, the gap structure of odd-parity pairing states $\ket{J,0}$ with odd $J$ is different. As discussed earlier in this section, see also Eq. \eqref{eq:mirrorsymmetry}, mirror symmetries force the pairing gap to vanish at points on the Fermi surface. In the present case, since the $\ket{J,0}$ have a continuous rotation symmetry about the $z$-axis, the point nodes are located on the $z$-axis, i.e., at $\pm \bK = k_{Fv}\hat z$, see Sec. \ref{ssec:rotaxis}. To demonstrate this in more detail, we rearrange the spinor components of $\Psi_{\bq \up,\down }$ in Eq. \eqref{eq:Psiq} and define $\Psi_{\bq 1,2 } $ as
\be
\Psi_{\bq 1 } = \begin{pmatrix} f_{\bq 1 } \\  \epsilon f^{\dagger \trans}_{-\bq 2  } \end{pmatrix}, \quad \Psi_{\bq 2 } = \begin{pmatrix}  f_{\bq 2 } \\  \epsilon f^{\dagger \trans}_{-\bq 1  } \end{pmatrix}. \label{eq:Psi12}
\ee
The low-energy pairing hamiltonian near $\pm \bK$, defined in Eq. \eqref{eq:LEpairinghamSodd}, can be expressed as
\begin{multline}
H_{\Delta} \simeq \frac12 \sum_{\bq}  \Big[ \Psi^\dagger_{\bq 1 } (\Delta_{\bq} \tau_++\Delta^\dagger_{\bq} \tau_- )  \Psi_{\bq 1 } \\ - \Psi^\dagger_{\bq 2 } (\Delta_{-\bq} \tau_++\Delta^\dagger_{-\bq} \tau_- ) \Psi_{\bq 2 } \Big], \label{eq:H12}
\end{multline}
where we recall that $\tau_{\pm} = (\tau_x \pm i\tau_y)/2$, and $ \Delta_{\bq}$ is given by
\be
 \Delta_{\bq}  =  \frac{1}{\sqrt{2}} \Delta_{\bq 0} s_z - \Delta_{\bq +}s_+   +  \Delta_{\bq -} s_- .  \label{eq:deltadirac}
\ee
Here, as before, $s_z$ and $s_{\pm} = (s_x \pm i  s_y)/2$ are Pauli matrices in pseudospin space. Rotation symmetry and mirror symmetry impose constraints on the three functions $\{ \Delta_{\bq+},\Delta_{\bq 0 },\Delta_{\bq -}\}$. [Note that the action of mirror symmetries on the pseudospin degrees of freedom was determined in Eq. \eqref{eq:mirrorsymmetry}.] In particular, under rotations $C_{\theta z}$ by an angle $\theta$ the spin matrices transform as $C_{\theta z}\,:\,s_\pm \rightarrow e^{\pm 3i\theta}s_\pm$; $s_z$ is left invariant. As a result, to lowest linear order in $\bq$ one finds $\Delta_{\bq 0 }=0$ and $\Delta_{\bq \pm}  \propto \Delta (q_x \mp i q_y)^3$, where $\Delta$ is the pairing amplitude. This not only shows that the pairing gap vanishes on the rotation $z$-axis, but also proves that the quasiparticle dispersion is \emph{cubic} in momenta $q_{x,y}$ in directions orthogonal to the rotation axis. 

Similarly, one may consider Eq. \eqref{eq:deltadirac} for the pseudospin $\pm \frac12$ Fermi surface, in which case the Hamiltonian \eqref{eq:H12} should be expressed in terms of spinors $\Phi_{\bq 1,2 } $ defined in analogy with Eq. \eqref{eq:Psi12}. One then finds that $\Delta_{\bq 0 }=0$ and $\Delta_{\bq \pm}  \propto \Delta (q_x \mp i q_y)$, implying that the quasiparticle dispersion is linear in all directions away from $\pm \bK$.

We can express this in terms of a Hamiltonian $\mathcal{H}_{\bq}$ for the low-energy quasiparticles at $\pm \bK$. Introducing a set of Pauli matrices $\sigma_{x,y,z}$ for the nodal degree of freedom, i.e., $\sigma_z=\pm 1$ for $\pm \bK$, and including the normal state contribution of Eq. \eqref{eq:Hqvc}, one finds that $\mathcal{H}_{\bq}$, in the basis of $\Psi_{\bq \up,\down }$  [defined in Eq. \eqref{eq:Psiq}], takes the form
\be
\mathcal{H}_{\bq} =v_{Fv}q_z\sigma_z + \Delta \sigma_x[ (q^3_++q_-^3) \tau_x +i(q^3_+-q_-^3) s_z\tau_y ], \label{eq:HamDiracM=0-3/2}
\ee
where $q_\pm = (q_x \pm i q_y)$. A Hamiltonian of this form describes Dirac quasiparticles with cubic dispersion. We thus draw the important conclusion that the pseudospin $\pm \frac32$ gap structure of odd-parity pairing states $\ket{J,0}$ with odd $J$ is given by triple Dirac points on the rotation axis.


For the pseudospin $\pm \frac12$ states, on the other hand, the Hamiltonian for the low-energy gapless quasiparticles is obtained as
\be
\mathcal{H}_{\bq} =v_{Fv}q_z\sigma_z + v_\Delta \sigma_x( q_x \tau_x - q_y s_z\tau_y ) , \label{eq:HamDiracM=0-1/2}
\ee
where $v_\Delta$ is an effective velocity in the $x,y$ directions, set by the pairing strength. Hamiltonian \eqref{eq:HamDiracM=0-1/2} shows that the low-energy quasiparticles are governed by a Dirac equation with linear dispersion.

\begin{table}[t]
\centering
\begin{ruledtabular}
\begin{tabular}{ccc}
Parity & $J$ & Gap structure topology  \\ [4pt]
\hline  
Even & Even & Line nodes  \\
 & Odd & Fully gapless, no pairing  \\
 \hline
Odd & Even  & Fully gapped, class DIII topological \\
 & Odd  & Point nodes, Dirac superconductor
\end{tabular}
\end{ruledtabular}
 \caption{{\bf Gap structure of $\ket{J,0}$ pairing states.} Gap structures topology of multicomponent time-reversal invariant pairing states with $M=0$ axial angular momentum. Odd-parity pairing states with odd $J$ have bulk nodes with gapless quasiparticles described by the Hamiltonian of Eq. \eqref{eq:HamDiracM=0}. Notably, the odd-parity states with even $J$ are fully gapped topological superconductors in class DIII, which break rotational symmetry and have subsidiary nematic order. Note that the results summarizes in this table are not specific to the valence band Fermi surface, but also hold for a conduction band Fermi surface (i.e., electron doping). 
 }
\label{tab:gapM=0}
\end{table}

As an example of an odd-parity pairing state $\ket{J,0}$ which gives rise to Dirac points on the rotation axis one may consider the state $\ket{3,0}$ given by
\be
\ket{3,0}=\mathsf{J}_{30} = \sqrt{\frac{1}{2}} ( Y_{11}\mathsf{S}_{3,-1}- Y_{1,- 1}\mathsf{S}_{31} ).  \label{eq:example30}
\ee
Here, $Y_{1,\pm 1} = Y_{1,\pm 1}(\hat \bk)$ are spherical harmonics. Since $Y_{1,\pm 1} \propto (k_x \pm ik_y)$ it is clear that the quasiparticle spectrum remains gapless on the rotation $z$-axis. 

Similarly, one may consider simple examples of odd-parity pairing states with a full gap on the Fermi surface. The pairing state $\ket{2,0}$, for instance, takes the form 
\be
\ket{2,0}=\mathsf{ J}_{20} = \frac{1}{\sqrt{6}}(Y_{11}\mathsf{ S}_{1,-1}+2Y_{10}\mathsf{ S}_{10}+Y_{1,-1}\mathsf{ S}_{11}).   \label{eq:example20}
\ee
This can be rewritten as $\ket{2,0}\propto 2k_zS_z -k_xS_x-k_yS_y $, showing that it is one of the five components of a rank-2 traceless symmetric tensor. In contrast to Eq. \eqref{eq:example30}, the pairing does not vanish along the rotation $z$-axis. It is straightforward to verify that Eq. \eqref{eq:example20} gives rise to a full pairing gap on the Fermi surface. Note that the pairing of Eq. \eqref{eq:example20} derives from the coupling of orbital angular momentum $L=1$ and spin angular momentum $S=1$. The same pairing state may, for instance, be realized in the $(L,S;J)=(1,3;2)$ channel and takes the form
\be
\mathsf{ J}_{20} = \sqrt{\frac{2}{7}}(Y_{11}\mathsf{ S}_{3,-1}-\sqrt{\frac{3}{2}}Y_{10}\mathsf{ S}_{30}+Y_{1,- 1}\mathsf{ S}_{31}) .\label{eq:example20b}
\ee

We now come to even-parity $\ket{J,0}$ pairing states. Once more, mirror reflections prove to be important. The even-parity pairing states are odd (even) under mirror reflections in planes perpendicular to the $xy$-plane when $J$ is odd (even). First, consider odd $J$. States with odd $J$ are odd under mirror reflections and this forces the gap function to vanish on any of these mirror planes. (Note that even-parity pairing states are pseudospin singlets.) Since the states $\ket{J,0}$ have a continuous rotation symmetry about the $z$-axis, this implies that the pairing gap vanishes on the entire Fermi surface. As a result, the even-parity pairing states $\ket{J,0}$ with odd $J$ remain fully gapless. 

Finally, the even-parity states $\ket{J,0}$ with even $J$ are mirror symmetric and generically have line nodes. The latter simply follows from the fact that the pseudospin-singlet gap function must have sign changes on the Fermi surface.

These results are summarized in Table \ref{tab:gapM=0}. It is important to point out that these results are independent of whether a valence band or conduction band Fermi surface is considered. In particular, Eqs. \eqref{eq:H12} and \eqref{eq:deltadirac}, and subsequent analysis, remain valid when applied to a conduction band Fermi surface.

\subsection{Low-energy gap structure of chiral pairing states \label{ssec:gapchiral}}

We proceed to another class of stationary states: the chiral pairing states $\ket{J,M}$ with nonzero $M$. These states break time-reversal symmetry, and since $\ket{J,M}$ and $\ket{J,-M}$ are time-reversed partners, we only have to consider $M \ge 1$. The key symmetry property of the chiral states is their eigenvalue of rotation about the quantization axis, which depends on the axial angular momentum $M$. As discussed in Sec. \ref{ssec:symmetries}, we can take the $z$-axis as the rotation axis. Then, focusing on $\pm \bK =\pm k_{Fv}\hat z$, in this section the aim is to derive the constraints on the gap functions $\Delta_{\bq s}$ and $\{ \Delta_{\bq+},\Delta_{\bq 0 },\Delta_{\bq -} \}$ of Eqs. \eqref{eq:LEpairinghamSeven} and \eqref{eq:LEpairinghamSodd} imposed by rotation symmetry. 

Rotation symmetry imposes the constraint that the orbital angular momentum of the low-energy gap functions $\Delta_{\bq s}$ and $\{ \Delta_{\bq+},\Delta_{\bq 0 },\Delta_{\bq -} \}$ must match the axial angular momentum $M$. Here, we anticipate differences for a valence band and conduction band Fermi surface. The pseudospin-triplet operators $F^\dagger_{t \pm}$ of Eq. \eqref{eq:Ftriplet} carry angular momentum $\pm 3$, whereas in the case of a conduction band Fermi surface they carry angular momentum $\pm 1$, affecting the matching conditions. 

As a result of time-reversal symmetry breaking, the chiral pairing states are generically spin-selective have nonzero $\delta$ in Eqs. \eqref{eq:LEpairinghamSeven} and \eqref{eq:LEpairinghamSodd}. Within the class of spin-selective chiral pairing states we can formulate a more precise constraint on $\delta$ by considering the full pairing potential $\Delta_\bk$ of Eq. \eqref{eq:deltak} at $\pm \bK$. If $\Delta_\bk$ vanishes at $\pm \bK$, i.e., $\Delta_{\pm \bK}=\mathsf{J}_{JM_J}(\pm \hat{\bK}) =0 $, then $\delta$ must be zero. Since only $Y_{L0}(\pm \bK)\neq 0$, the pairing potential $\Delta_{\pm \bK}  $ must vanish whenever the corresponding Clebsch-Gordan coefficient vanishes, i.e., 
\be
\bracket{M_L=0,M_S=M}{JM}=0.
\ee
For the $L=1$ multiplets this only occurs for $(L,S;J)=(1,1;2)$ when $M=2$.

\tocless\subsubsection{Pseudospin-singlet Hamiltonian from symmetry \label{sssec:singlet}}

The pseudospin-singlet pairing Hamiltonian defined in Eq. \eqref{eq:LEpairinghamSeven} can be combined with normal state Hamiltonian Eq. \eqref{eq:Hqvc} to obtain the full Hamiltonian of the low-energy quasiparticle degrees of freedom. Rather than $\Psi_{\bq \up,\down }$, it is convenient to rearrange the operators and form the spinors $\Psi_{\bq \pm } $ given by
\be
\Psi_{\bq + } = \begin{pmatrix} f_{\bq 1 \up} \\  f_{\bq 2  \up} \\  f^\dagger_{-\bq 1  \down}  \\  f^\dagger_{-\bq 2  \down}  \end{pmatrix}, \quad \Psi_{\bq - } = \begin{pmatrix} f_{\bq 1 \down} \\  f_{\bq 2  \down} \\  f^\dagger_{-\bq 1  \up}  \\  f^\dagger_{-\bq 2  \up}  \end{pmatrix}. \label{eq:Psipm}
\ee
The full Hamiltonian is block diagonal in this basis, i.e., $\mathcal{H}_{\pm} =\frac12 \sum_{\bq}  \Psi^\dagger_{\bq \pm } \mathcal{H}_{\bq \pm} \Psi_{\bq \pm }$, with the Hamiltonian matrices $\mathcal{H}_{\bq \pm}$ are given by
\be
\mathcal{H}_{\bq\pm}  = \begin{pmatrix} \varepsilon^{v}_{\bq} \pm \delta  & 0  & 0   & \pm \Delta_{\bq s} \\  0 & \varepsilon^{v}_{-\bq} \pm \delta & \pm \Delta_{-\bq s} & 0 \\ 0  &  \pm \Delta^*_{-\bq s}& - \varepsilon^{v}_{-\bq}  \pm \delta& 0  \\  \pm \Delta^*_{-\bq s}& 0&0 &  -\varepsilon^{v}_{\bq}  \pm \delta \end{pmatrix}  \label{eq:Hpm}
\ee
The constraint on the gap function $\Delta_{\bq s} $ is that the angular momentum of the pseudospin-singlet pairing matches $M$. More specifically, the gap function must be of the form
\be
\Delta_{\bq s}  \propto \Delta (q_x + i q_y)^{M} .  \label{eq:singletmatching}
\ee
The Hamiltonian of Eq. \eqref{eq:Hpm} can be recast using Pauli matrices $\sigma_z = \pm 1$ for the nodal degree of freedom $\pm \bK$. Equation \eqref{eq:singletmatching} shows that $\bdelta_\bq$ is either even or odd under $\bq \to -\bq$. When the gap function is even, Eq. \eqref{eq:Hpm} can be expressed as
\be
\mathcal{H}_{\bq \pm}  = v_{Fv}q_z\sigma_z \pm \sigma_x( \text{Re}\,\Delta_{\bq s} \tau_x - \text{Im}\,\Delta_{\bq s} \tau_y ) \pm \delta  , \label{eq:Hpm_even}
\ee
and when it is odd we find
\be
\mathcal{H}_{\bq \pm}  = v_{Fv}q_z\sigma_z \mp \sigma_y(   \text{Im}\,\Delta_{\bq s} \tau_x +\text{Re}\,\Delta_{\bq s} \tau_y ) \pm \delta  .\label{eq:Hpm_odd}
\ee
The spectrum takes the same general form on both cases. We find the four energy eigenvalues
\be
E^{\pm\pm}_\bq  = \pm \delta \pm \sqrt{(v_{Fv}q_z)^2 + |\Delta_{\bq s}|^2}.  \label{eq:Esinglet}
\ee
The most significant feature of these solutions is that they generically describe two nodes of codimension 1, one enclosing $\bK$ and one enclosing $-\bK$. These nodes are defines by the solutions of the equation $|\delta| = \sqrt{(v_{Fv}q_z)^2 + |\Delta_\bq|^2}$. Bogoliubov Fermi surfaces of this type were first described in Ref. \cite{agterberg17}, which noted that they may be viewed as inflated point nodes. This interpretation naturally follows from the picture presented here. That Fermi surfaces may be topologically stable features of a gap structure follows from a topological $\mathbb{Z}_2$ invariant associated with nodes of codimension 1 in even-parity time-reversal symmetry broken superconductors \cite{kobayashi14,zhao16}.

Additional Bogoliubov Fermi surfaces generically occur on the Fermi surface equator, i.e., in the vicinity $k_z=0$, of even-parity chiral pairings with odd $M$. For odd $M$, $\ket{J,M}$ is odd under twofold rotation about the $z$-axis. Since we have assumed even parity, $\ket{J,M}$ is also odd under a mirror reflection in the $xy$-plane. This would imply a line node on the equator, however, since the pairing is spin-selective these line nodes generically will be inflated to nodes of codimension 1.

\tocless\subsubsection{Pseudospin-triplet Hamiltonian from symmetry \label{sssec:triplet}}

\begin{table}[t]
\centering
\begin{ruledtabular}
\begin{tabular}{ccccc}
$M$ & \multicolumn{2}{c}{Pseudospin $\pm\frac32$} & \multicolumn{2}{c}{Pseudospin $\pm\frac12$}  \\ 
 & $\Delta_{\bq +}$ &  $\Delta_{\bq -}$ &  $\Delta_{\bq +}$ &$\Delta_{\bq -}$   \\ [4pt]
\hline  
1  &   $\propto q^2_{-}$  &  $\propto q^4_{+}$    &  $\propto 1$    &  $\propto q^2_{+}$    \\
2  &   $\propto q_{-}$  &  $\propto q^5_{+}$    &  $\propto q_+$    &  $\propto q^3_{+}$    \\
3  &   $\propto 1 $  &  $\propto q^6_{+}$    &  $\propto q^2_+$    &  $\propto q^4_{+}$    \\
\end{tabular}
\end{ruledtabular}
 \caption{{\bf Gap functions of odd-parity $\ket{J,M}$ pairing states.} Leading order expansions of the gap functions $\Delta_{\bq +}$ and $\Delta_{\bq -}$ for chiral pairing states with $M=1,2,3$ according to Eqs. \eqref{eq:deltaqpmexpand} and \eqref{eq:tripletmatching} (and their equivalents for $\Delta_{\bq -}$). We have defined $q_\pm = q_x \pm i q_y$. Also shown are the leading order expansions of the gap functions for the case of a pseudospin $\pm\frac12$ Fermi surface, i.e., the conduction band in our case. 
 }
\label{tab:gapM}
\end{table}

Pseudospin-triplet pairing must be of odd-parity type, and therefore nodes of codimension 1 (i.e., surfaces) are not topologically stable \cite{kobayashi14,zhao16}. Since chiral pairing states are generically spin-selective, the effective pseudopsin-splitting $\delta$ is nonzero, implying that point nodes on the rotation axis (if they exist) are non-degenerate. This may be compared to non-degenerate nodal degeneracies in ferromagnetic superconductors, where the Zeeman splitting originates from ferromagnetic order rather than spin-selective pairing \cite{sau12}. 

The splitting of pseudospin-$\up$ and $\down$ implies that, in order to determine the symmetry-mandated low-energy dispersion of quasiparticles on the rotation $z$-axis, we need to examine the gap functions $\Delta_{\bq +}$ and $\Delta_{\bq -}$, since these correspond to $\up\up$- and $\down\down$-pairing. In the spirit of Refs. \onlinecite{fang12,kozii16,fang17} one finds that constraints derived from rotational symmetry (i.e., the angular momentum quantum numbers) fully determine the form of $\Delta_{\bq +}$ and $\Delta_{\bq -}$. We demonstrate this by considering $\Delta_{\bq +}$. The gap function $\Delta_{\bq +}$ can be expanded in momenta $q_x,q_y$ perpendicular to the rotation axis as 
\be
\Delta_{\bq +} = \sum_{\ell,\ell '} A_{\ell \ell '} (q_x+iq_y)^{\ell }(q_x-iq_y)^{\ell '},  \label{eq:deltaqpmexpand}
\ee
where $\ell'$ and $\ell'$ are nonnegative integers defining orbital axial angular momentum quantum numbers; $A_{\ell \ell'}$ are coefficients. In terms of the quantum numbers $\ell, \ell'$ the orbital angular momentum of $\Delta_{\bq +} $ is given by $\ell - \ell'$. Furthermore, in the case of the (valence band) pseudospin $\pm \frac32$ states the pseudospin angular momentum of $\up\up$-pairing is $\frac32+\frac32=3$. (The latter equals $\frac12+\frac12=1$ for the conduction band $\pm \frac12$ pseudospin states.) Since the total axial angular momentum of the pairing state $\ket{J,M}$ is $M$, the sum of orbital and pseudospin angular momentum must be equal to $M$, and we thus arrive at the matching condition
\be
M= 3 + \ell - \ell' .  \label{eq:tripletmatching}
\ee
Since in the expansion of Eq. \eqref{eq:deltaqpmexpand} we are interested in the lowest order terms, we only consider solutions of \eqref{eq:tripletmatching} for which either $\ell$ or $\ell'$ is zero. For instance, in the case when $M=2$, the matching condition gives $(\ell , \ell')=(0,1)$. Clearly, a matching condition similar to Eq. \eqref{eq:tripletmatching} exists for $\Delta_{\bq -}$, in which case the pseudospin angular momentum is $-3$. Applying these matching conditions to the cases $M=1,2,3$, we arrive at Table \ref{tab:gapM}. 

An analogous analysis can be straightforwardly performed for a pseudospin $\pm \frac12$ conduction band Fermi surface, in which case the pseudospin-triplet operators carry angular momentum $\pm 1$ and one should replace $3$ with $1$ in \eqref{eq:tripletmatching}. The low-energy behavior of pseudospin $\pm \frac12$ gap functions $\Delta_{\bq \pm }$ is summarized in Table \ref{tab:gapM} as well.

Table \ref{tab:gapM} shows that special cases arise when $M=3$ (for pseudospin $\frac32$) and $M=1$ (for pseudospin $\frac12$) , i.e., when the angular momentum of the pairing state matches the angular momentum of the operator for pseudospin-$\up\up$ pairing. In this case, the pseudospin-$\up$ quasiparticles can pair at $\pm \bK$ and develop a pairing gap. The pseudospin-$\down$ quasiparticles must remain gapless, however. Therefore, when the angular momentum of the pairing state matches the angular momentum of the Cooper pair $ f^\dagger_{1 \mu}f^\dagger_{ 2 \mu}$ ($\mu=\up$ or $\down$), one of the two nodes along the rotation axis will be gapped out, leaving a single 3D Majorana fermion behind.

The low-energy spectral properties of chiral pairing states $\ket{J,M\neq 0}$ summarized in Table \ref{tab:gapM} have been established based on symmetry arguments which take into account the pseudospin splitting $\delta$ implicitly. The low-energy quasiparticle gap structure of odd-parity chiral pairing states may also be obtained from an explicit calculation based on the Hamiltonian of Eq. \eqref{eq:LEpairinghamSodd}. Specifically, using the spinors defined in Eq. \eqref{eq:Psi12}, the pairing Hamiltonian $H_{\Delta}$ of Eq. \eqref{eq:LEpairinghamSodd} (which explicitly includes the pseudospin splitting proportional to $\delta$) is expressed as 
\be
H_{\Delta} \simeq \frac12 \sum_{\bq} \Psi^\dagger_{\bq 1 } (\Delta_{\bq} \tau_++\Delta^\dagger_{\bq} \tau_- )  \Psi_{\bq 1 } + \delta\, \Psi^\dagger_{\bq 1 }s_z \Psi_{\bq 1 }  ,\label{eq:tripletham}
\ee
where have suppressed the contribution from $\Psi_{\bq 2 } $ since all spectral information is contained in \eqref{eq:tripletham}. As in Eq. \eqref{eq:H12}, the pairing potential $\Delta_{\bq}$ contains the three gap functions $\{ \Delta_{\bq+},\Delta_{\bq 0 },\Delta_{\bq -}\}$. These gap functions as well as $\delta$ can be determined using the perturbative approach detailed in Appendix \ref{app:qpproject}. 


Upon including the normal state part of the Hamiltonian, to the lowest order given by $\varepsilon^v_\bq \tau_z$,  we obtain the four branches $E^{\pm\pm}_\bq$ of the quasiparticle spectrum as
\begin{multline}
E^{\pm\pm}_\bq  = \pm \left[(\varepsilon^v_\bq)^2 +\delta^2 +\frac12|\Delta_{\bq +} |^2 + \frac12|\Delta_{\bq -} |^2 \right. \\ 
\left. +\frac12|\Delta_{\bq 0} |^2 \pm \frac12\Lambda_\bq  \right]^{1/2}.  \label{eq:Echiral}
\end{multline}
where $ \Lambda_\bq $ is defined as
\begin{multline}
\Lambda_\bq  = \Big[ (|\Delta_{\bq +} |^2 - |\Delta_{\bq -} |^2)^2+2(\Delta^*_{\bq +}\Delta^*_{\bq -}\Delta^2_{\bq 0}+\text{c.c.})   \\
+ 2(|\Delta_{\bq +} |^2 + |\Delta_{\bq -} |^2)|\Delta_{\bq 0} |^2 + 8\delta^2 |\Delta_{\bq 0}|^2 +8(\varepsilon^v_\bq\delta)^2\\
8\varepsilon^v_\bq\delta (|\Delta_{\bq +} |^2 - |\Delta_{\bq -} |^2) \Big]^{1/2} .  \label{eq:Lambda}
\end{multline}
It is straightforward to establish that $\Lambda_\bq $ is only nonzero for chiral states and must be zero when time-reversal symmetry is present. To see this, note that time-reversal symmetry requires $\Delta^*_{\bq 0} = \Delta_{\bq 0}$, $\Delta^*_{\bq +} = -\Delta_{\bq -}$, $\Delta^*_{\bq -} = -\Delta_{\bq +}$, and, as discussed above, $\delta=0$. It then simply follows that $\Lambda_\bq=0$ in this case.

Even though Eqs. \eqref{eq:Echiral} and \eqref{eq:Lambda} appear rather complicated, they describe low-energy gap structures whose essential properties have been rigorously determined from symmetry arguments, and are given by Table \ref{tab:gapM}. What Eqs. \eqref{eq:Echiral} and \eqref{eq:Lambda} nevertheless serve to illustrate is the importance of the energy scale set by $\delta$. In particular, as mentioned in Sec. \ref{ssec:rotaxis}, $\delta$ describes a pairing-induced splitting of pseudospin states proportional to $ |\Delta|^2/\varepsilon^c_\bK$. Correspondingly, as may be checked directly from Eqs. \eqref{eq:Echiral} and \eqref{eq:Lambda}, the two point nodes are separated by a momentum $\sim  |\Delta|^2/(v_{Fv} \varepsilon^c_\bK )$ along the $k_z$ axis. The emergence of this energy scale is important for the potential observation of the nodal structure  through thermodynamic probes, since it determines the temperature range over which the characteristic temperature dependence of thermodynamic quantities is accessible.

We conclude the discussion of odd-parity chiral pairing states by illustrating the general considerations with simple examples. We focus our attention on the chiral pairing states with $M=1,2,3$ in a $J=3$ channel with angular momentum quantum numbers $(L,S)=(1,3)$. The corresponding pairings $\mathsf{J}_{3M}$ are given by
\beq
\mathsf{J}_{31} &=&  \frac{1}{\sqrt{2}}(Y_{11}\mathsf{S}_{30} - \frac{1}{\sqrt{6}}Y_{10}\mathsf{S}_{31}- \sqrt{\frac{5}{6}}Y_{1,- 1}\mathsf{S}_{3 2}), \nonumber \\
\mathsf{J}_{32} &=&  \sqrt{\frac{5}{12}}Y_{11}\mathsf{S}_{31} - \frac{1}{\sqrt{3}}Y_{10}\mathsf{S}_{32} -\frac{1}{2}Y_{1,- 1}\mathsf{S}_{33}, \nonumber \\
\mathsf{J}_{33} &=& \frac{1}{2}( Y_{11}\mathsf{S}_{32}-\sqrt{3}Y_{10}\mathsf{S}_{33}), \label{eq:examplechiral}
\eeq
where we have suppressed the momentum dependence of the spherical harmonics. These pairings generate the low-energy nodal structures listed in Table \ref{tab:gapM}. To gain a better understanding of these example pairings, consider the terms proportional to $Y_{10} \sim k_z$, which are nonzero on the rotation $z$-axis. In case of the $\mathsf{J}_{31}$ pairing with $M=1$, the spin matrix $\mathsf{S}_{31}$ connects states which differ by one unit of angular momentum. Therefore, $\mathsf{S}_{31}$ does not directly connect the pseudospin-$\frac32$ states, which have relative angular momentum $3$, but does connect the pseudospin-$\frac12$ states. As a result, both pseudospin-$ \frac32$ species remain gapless, whereas the pseudospin $+\frac12$ states can pair, leaving only the pseudospin $-\frac12$ gapless. This qualitatively explains the first row of Table \ref{tab:gapM}. A similar argument can be made in case of $\mathsf{J}_{33}$ pairing: the spin matrix $\mathsf{S}_{33}$ connects the pseudospin-$\frac32$ states, such that a pseudospin-$\up\up$ pairing can form on the rotation axis. This corresponds to the aforementioned special case of $M=3$; see also Table \ref{tab:gapM}. In contrast, the matrix $\mathsf{S}_{32}$ does not directly connect any of the states within a pseudospin sector. In all these three cases, the terms proportional to $Y_{10}$ are responsible for finite $\delta$, giving rise to the pseudospin splitting.

\begin{table*}[t]
\centering
\begin{ruledtabular}
\begin{tabular}{cclc}
Symmetry class &  $\Theta$     &  Gap structure and Topology  & Exists in channel $J$  \\ [4pt]
\hline
\multicolumn{4}{c}{\emph{(a)} Odd parity} \\[4pt]
$O$ & Yes    & Topological SC in class DIII ($J=4$) / Dirac superconductor ($J=3$) & $3,4$ \\[2pt]
$T$ & No  & Majorana fermions at $\bk_F$ along $(111)$ and equivalent directions  & $2,4$ \\[2pt]
 $D_8$ & Yes & Dirac superconductor with linear or cubic dispersion   & $4$ \\[2pt]
 $D_6$ & Yes & Dirac superconductor with linear or quadratic dispersion   & $3$ \\[2pt]
$D_4$ & Yes   & Dirac superconductor with linear dispersion  & $2$  \\[2pt]
$D_3$ & No   & Majorana fermions or pairing gap along the threefold axis  & $3$  \\[2pt]
\multicolumn{4}{c}{\emph{(b)} Even parity} \\[4pt]
$T$ & No  & Bogoliubov Fermi surfaces and line nodes   & $2$ \\[2pt]
$D_4$ & Yes   & Line nodes  & $2$  \\[2pt]
\end{tabular}
\end{ruledtabular}
 \caption{ {\bf Pairing states with discrete symmetry.} Table summarizing the considered gap structures of pairing states with discrete symmetry. The first column lists the symmetry classes discussed in this section; the presence or absence of time-reversal symmetry for states with given symmetry is indicated. The final column indicates in which pairing channel, labeled by $J$, pairing states with given symmetry exist. We explicitly distinguish even- and odd-parity pairing. In this table we only focus on pairing channels $J$ which can be formed up to $p$-wave order, i.e., $L=1$, see Table \ref{tab:classification}. 
 }
\label{tab:classification2}
\end{table*}

\subsection{Gap structures of pairing states with discrete symmetry \label{ssec:gapdiscrete}}

In the final part of this section, we consider gap structures of pairing states with discrete symmetry. As mentioned in Sec. \ref{ssec:symmetries}, even when the normal state has full $SO(3)$ rotational symmetry, there can be---and typically will be---stationary states of the free energy with discrete spatial symmetry. In Sec. \ref{ssec:GLexamples} we have discussed a number of examples of such pairing states, focusing in particular on the inert states. Here, we examine their gap structures. 

In order to do so, it is necessary to comment on the precise structure of the isotropy groups of these pairing states. We have briefly mentioned the definition of isotropy groups in Sec. \ref{ssec:GLexamples}; they are subgroups of the full symmetry group $G$ which leave the state invariant. Importantly, this implies that elements of the isotropy group may be composites of spatial transformations and $U(1)$ gauge factors. (We have implicitly made use of this in the case of the continuous isotropy groups of the $\ket{J,M}$ states.) In the case of discrete isotropy groups such as $O$, $T$, and $D_n$, it is particularly important to properly account for phase factors associated with spatial symmetries. Consider, for instance, the pairing state $\ket{2,2}+\ket{2,-2}$ of a $J=2$ superconductor, which has $D_4$ symmetry. The two generators of the isotropy group $D_4$ are given by $\{ e^{i\pi}C_{4z},C_{2x} \}$, showing that the fourfold rotation leaves the state invariant only in combination with the phase factor $e^{i\pi}$. The significance of this for our purposes is that the precise structure of the isotropy group can depend on the total angular momentum of the superconductor. In particular, two pairing states with the same discrete symmetry may still have different gap structure due to a different realization of the isotropy group. 

We further note that the isotropy groups are taken to be subgroups of $U(1)\times SO(3)$. For superconductors, the symmetry of the pairing state under inversion, and more generally under improper rotations, is fixed by the parity of the pairing state. Therefore, we treat even- and odd-parity pairing separately.  

Our goal in this section is to illustrate our classification by focusing primarily on inert pairing states with $O$, $T$, and $D_n$ symmetry; one example of noninert pairing states will be explicitly discussed. Furthermore, we will consider specific pairing states with these symmetry groups up to total angular momentum $J=4$, the highest total angular momentum up to $p$-wave order, see Sec. \ref{sec:pairingclass}. Generalization to higher angular momentum channels is straightforward and will be mentioned where appropriate.  

\tocless\subsubsection{Pairing states with octahedral $O$ symmetry \label{sssec:O}}

First, we consider pairing states with isotropy group $O$, i.e., the group of all rotations which leave the octahedron and the cube invariant. Two examples of such states are given by [see Eqs. \eqref{eq:J3O}  and \eqref{eq:J4O}]
\beq
\ket{\bdelta_{O}}_{3} &=&  \ket{3,2} - \ket{3,-2}, \label{eq:octa2} \\
\ket{\bdelta_{O}}_4 &=& \sqrt{5}\ket{4,4}  + \sqrt{14}\ket{4,0}+\sqrt{5} \ket{4,-4},\label{eq:octa4}
\eeq
where we have indicated the angular momentum $J$ as $\ket{\bdelta_{O}}_J$. (Here, we will not be concerned with the normalization of these states.) The two generators of the isotropy groups are given by $\{e^{i\pi}C_{4z},e^{i\pi}C_{2,z+x} \}$ and $\{C_{4z},C_{2,z+x} \}$, respectively, where $C_{2,z+x}$ is a twofold rotation about the $(101)$ axis. We further observe that both states are time-reversal invariant pairing states, and are thus spin-degenerate. 

The state $\ket{\bdelta_{O}}_4$ is interesting, since it is invariant under all rotation symmetries of the cube. This implies in particular that odd-parity pairing states $\ket{\bdelta_{O}}_4$ do not possess any mirror symmetry. Based on the discussion of mirror symmetries presented in the introductory part of this section, and in light of similar considerations in Sec. \ref{ssec:gapnematic}, we conclude that odd-parity $\ket{\bdelta_{O}}_4$ states realize fully gapped topological superconductors in class DIII. Intriguingly, whereas the nematic topological superconductors described in Sec. \ref{ssec:gapnematic} have a nonzero quadrupole moment [in the sense of Eq. \eqref{eq:subsorder}], the symmetry of the $\ket{\bdelta_{O}}_4$ states does not allow a quadrupole moment. In fact, the highest nonzero multipole moment (i.e., subsidiary order) is a hexadecapole moment. 

The even-parity $\ket{\bdelta_{O}}_4$ states clearly do have mirror symmetry; the planes perpendicular to any of the twofold axes are mirror planes. In the even-parity case this does not mandate degeneracies and this generically leads to a full pairing gap for even-parity $\ket{\bdelta_{O}}_4$ states. These gapped even-parity superconductors are topologically trivial.

We turn to the $J=3$ states $\ket{\bdelta_{O}}_3$. Due to the phase factor associated with the twofold rotations the odd-parity $\ket{\bdelta_{O}}_3$ states are invariant under mirror reflection in planes perpendicular to the $(110) $ axis (and equivalent axes). In contrast, the even-parity $\ket{\bdelta_{O}}_3$ states are odd under mirror reflection in planes perpendicular to the $(110) $ axis (and equivalent axes). For the even-parity states this implies line nodes on the Fermi surface. Instead, for the odd-parity pairing states the mirror symmetries, in combination with time-reversal invariance, leads to point nodes on the Fermi surface, located along the $(001) $ as well as the $(111) $ directions (and all equivalent directions). 

To establish the dispersion of the low-energy gapless quasiparticles at the nodes, we proceed in the same way as in Sec. \ref{ssec:gapnematic}. We first treat the fourfold axis along $(001) $, i.e., the $z$-direction. The pseudospin-triplet pairing $ \Delta_{\bq}$ matrix was defined in Eqs. \eqref{eq:H12} and \eqref{eq:deltadirac}, and takes the form
\be
\Delta_{\bq}  =  \frac{1}{\sqrt{2}} \Delta_{\bq 0} s_z - \Delta_{\bq +}s_+   +  \Delta_{\bq -} s_- .  \label{eq:deltadirac2}
\ee
Under fourfold rotation one has $C_{4z}\,:\,s_\pm \rightarrow e^{\pm 3i\pi/2}s_\pm$; $s_z$ is invariant. Since the pairing must be odd under fourfold rotation one finds
\beq
\Delta_{\bq 0} &\propto &A_+(q_x +i q_y)^2+A_-(q_x -i q_y)^2,  \nonumber \\
\Delta_{\bq \pm } &\propto& B_{\pm}(q_x \mp i q_y),
\eeq
where $A_{\pm }$ and $B_{\pm }$ are expansion coefficients. One of the mirror planes is perpendicular to $(110) $ and, invoking the same arguments which led to Eq. \eqref{eq:mirrorsymmetry}, we find that the mirror operation acts on the pseudospin states as $O_{M_{(110)}} = -i \hat n \cdot \bs$, where $\hat n = (1,1,0)^\trans/\sqrt{2}$. (See also Appendix \ref{app:projection}.) Mirror symmetry then leads to the requirements $A_+=A_-=A$ and  $B_+=B_-=iB$, where we have also used time-reversal symmetry in the latter. The pairing thus takes the form  
\be
\Delta_{\bq}  =  \frac{A}{\sqrt{2}} (q^2_x-q^2_y) s_z +B(q_xs_y-q_ys_x),  \label{eq:deltadirac2}
\ee
which, in combination with the normal state contribution given in Eq. \eqref{eq:Enormalstate}, gives rise to a Dirac Hamiltonian for the gapless low-energy quasiparticles, with linear dispersion to lowest order [c.f. Eq. \eqref{eq:HamDiracM=0-1/2}]. We note that the analysis is similar for the case of a pseudospin-$ \frac12$ conduction band Fermi surface, giving rise to linear dispersion.

Now, consider the threefold axis along the $(111) $ direction. It is convenient to apply a global rotation to the pairing state $\ket{\bdelta_{O}}_3$ such that the threefold axis is oriented along the $z$-direction. Alternatively, one may view this as choosing local coordinates $(q'_x,q'_y)$ perpendicular to the rotation axis, see Fig. \ref{fig:figFS}. We furthermore imagine that the state has been rotated such that the mirror symmetry is given by $M_y \,:\, y \to -y$. The group of symmetries which leave the intersection of the threefold axis and the Fermi surface $\bK$ invariant is given by $C_{3v}$, i.e., the threefold rotations and three equivalent mirror reflections. Importantly, it follows from group theory that the symmetry group $C_{3v}$ does not protect degeneracies for pseudospin-$\frac32$ states. In the present case this implies that no point nodes exist along the $(111) $ direction for a pseudospin-$\frac32$ Fermi surface. In the odd-parity pairing state $\ket{\bdelta_{O}}_3$ the (valence band) $\pm\frac32$ Fermi surface only exhibits point nodes along the $(001)$ axis (and equivalent axes). 


This is indeed different for pseudospin-$ \frac12$ states: a pseudospin-$ \frac12$ Fermi surface exhibits points along $(111)$ direction. Using the threefold rotations and mirror symmetries (in the rotated basis) we obtain the low-energy pairing matrix $\Delta_{\bq}$ given by
\be
\Delta_{\bq}  =  \frac{iA}{\sqrt{2}} (q'^3_+-q'^3_-) s_z +iB(q'_-s_+-q'_+s_-),  \label{eq:D6dirac2}
\ee
where we have defined $q'_{\pm} = q'_x \pm iq'_y$. This defines another set of Dirac points, in addition to the Dirac points along the $(001)$ direction.

\tocless\subsubsection{Pairing states with tetrahedral $T$ symmetry \label{sssec:T}}

Next, we consider pairing states with tetrahedral symmetry. As compared to the octahedral states, these pairing states lack a fourfold rotation axis; two examples are given by
\beq
\ket{\bdelta_{T}}_2 &=&  \ket{2,2} +i\sqrt{2} \ket{2,0} +  \ket{2,-2}, \label{eq:tetra2} \\
\ket{\bdelta_{T}}_4 &=&  \sqrt{7}\ket{4,4} +2i\sqrt{3} \ket{4,2}-\sqrt{10}\ket{4,0} \nonumber \\
&& +2i\sqrt{3}\ket{4,-2} + \sqrt{7} \ket{4,-4}.\label{eq:tetra4}
\eeq
The two generators of the respective isotropy groups are given by $\{e^{\mp i2\pi/3}C_{3\hat n},C_{2z} \}$, where $-,+$ applies to $J=2,4$, and $\hat n = (1,1,1)^\trans/\sqrt{3}$ is a unit vector along the threefold axis. Importantly, the tetrahedral states break time-reversal symmetry but do not have a chirality [Eq. \eqref{eq:chirality}], which follows directly from Eqs. \eqref{eq:tetra2} and \eqref{eq:tetra4}. This is true for general tetrahedral pairing states. As a result, tetrahedral pairing is spin-selective. 

Given that the tetrahedral pairing states have a threefold axis along the $(111)$ and equivalent directions, we can invoke the arguments of Secs. \ref{ssec:rotaxis} and \ref{ssec:gapchiral} to study the low-energy gap structure at Fermi momenta $\pm \bK$ defined along the $(111)$ rotation axis. Consider odd-parity pairing first. As is the case for chiral pairing states (see Sec. \ref{ssec:gapchiral}) we can focus on the gap functions $\Delta_{\bq \pm } $ for pseudospin-$\up$ and -$\down$ pairing. The pseudospin-triplet operators have angular momentum $\pm 3$ and are therefore invariant under threefold rotations. Consequently, the orbital angular momentum of $\Delta_{\bq \pm } $ must match the rotation eigenvalue of the pairing state. This implies that both $\Delta_{\bq \pm } \propto (q_x -iq_y )$ for $\ket{\bdelta_{T}}_2$, and, similarly, $\Delta_{\bq \pm } \propto (q_x +iq_y )$ for $\ket{\bdelta_{T}}_4$, giving rise to linearly dispersing Majorana fermions along the rotation axes in each pseudospin sector.  

In the case of a conduction band Fermi surface, i.e., when the pseudospin-triplet operators carry angular momentum $\pm 1$ and transform as $e^{\pm i2\pi/3}$ under threefold rotation, either $\Delta_{\bq + } $ or $\Delta_{\bq - } $ can acquire a constant non-$\bq$-dependent part. Therefore, only a single pseudospin species of Majorana fermions exists on the rotation axis: the Majorana fermions are fully spin-polarized. A realization of such Majorana fermions were theoretically found in the tetrahedral pairing state of $^3P_2$ superfluids \cite{mizushima17}. 

Majorana fermions on the threefold rotation axis are a generic property of tetrahedral pairing states, and the angular momentum of the paired electrons determines whether Majorana fermions of a single or both pseudospin species is present. 

Finally, we note that even-parity pairing states with tetrahedral symmetry will generically have $\mathbb{Z}_2$ Fermi surfaces  enclosing the Fermi momenta along the threefold rotation axes. This follows from the arguments presented in Sec. \ref{ssec:gapchiral}. In addition, since the even-parity pairing have a mirror plane orthogonal to the twofold rotation, they generically feature line nodal degeneracies as well.

\tocless\subsubsection{Pairing states with dihedral $D_n$ symmetry \label{sssec:Dn}}

At last, we turn to the class of dihedral pairing states with isotropy groups $D_n$. As a first example, consider the case $n=8$ and the pairing state $\ket{\bdelta_{D_8}}_4 = \ket{4,4} - \ket{4,-4} $ of Eq. \eqref{eq:J4D8}. This is a time-reversal invariant pairing state with isotropy group $\{e^{i\pi}C_{8z},e^{i\pi}C_{2x} \}$. For odd-parity pairing states, this implies four symmetry-related mirror planes (e.g., the $xz$ and $yz$ planes are both mirror planes). As a result, point nodes must appear along the eightfold axis, i.e., the $z$-axis. To obtain the low-energy gap structure of the point nodes at momenta $\pm \bK$ along rotation axis, we must require that $\Delta_{\bq}$ is odd under eightfold rotation and respects all mirror symmetries. Using that $s_\pm$ transform as $C_{nz}\,:\,s_\pm \rightarrow e^{\pm 6i\pi/n}s_\pm$ for pseudospin $\pm \frac32$ states, we find 
\be
\Delta_{\bq}  =  \frac{A}{\sqrt{2}} (q^4_++q^4_-) s_z +iB(q_+s_+-q_-s_-).  \label{eq:D8dirac1}
\ee
Instead, for pseudospin $\pm \frac12$ fermions transforming as $C_{nz}\,:\,s_\pm \rightarrow e^{\pm 2i\pi/n}s_\pm$ we find that the low-energy quasiparticle dispersion takes the form
\be
\Delta_{\bq}  =  \frac{A}{\sqrt{2}} (q^4_++q^4_-) s_z +iB(q^3_+s_+-q^3_-s_-).  \label{eq:D8dirac3}
\ee
As before, in both cases the significance lies in the second term. The low-energy gap structure describes quasiparticles with linear and \emph{cubic} Dirac dispersion, respectively. For the $\pm \frac12$ pseudospin states we thus find a another new type of low-energy quasiparticle: Dirac fermions with cubic dispersion. 

Next, consider $n=6$ with the pairing states are given by [see Eqs. \eqref{eq:J3D6} and \eqref{eq:J4D6}]
\beq
\ket{\bdelta_{D_6}}_3 &=&  \ket{3,3} + \ket{3,-3},\label{eq:D63} \\
\ket{\bdelta_{D_6}}_4 &=& \ket{4,3} - \ket{4,-3} .\label{eq:D64}
\eeq
The structure of the isotropy group is the same in both cases and given by $\{e^{i\pi}C_{6z},e^{i\pi}C_{2x} \}$. In addition, the two dihedral states are time-reversal invariant. The latter is true for all pairing states with $D_6$ symmetry. Clearly, referring earlier arguments, even-parity $\ket{\bdelta_{D_6}}_3$ and $\ket{\bdelta_{D_6}}_4$ pairing states are odd under certain mirror reflections and must therefore have line nodes. 

For the odd-parity states we make the following observations. The structure of the isotropy group generators implies that the odd-parity $\ket{\bdelta_{D_6}}$ states have a set of three vertical mirror planes given by the mirror operation $M_x \,:\, x \to -x$ and its two equivalents related by threefold rotation. In addition, the odd-parity states are invariant under mirror reflection in the $xy$ plane, since the isotropy group contains the element $e^{i\pi}C_{2z}$. These constraints have different implications for the pseudospin $\pm \frac32$ and $\pm \frac12$ Fermi surfaces. In particular, in the case of a pseudospin-$ \frac32$ Fermi surface, no points nodes are present along the sixfold rotation axis, i.e., the $(001)$ direction. This follows from the requirement that $\Delta_{\bq}$ must be invariant under the subgroup $C_{3v}$ and odd under the sixfold rotations, which is not sufficient to force $\Delta_{\bq=0}$ to vanish for $\pm \frac32$ doublets. In contrast, a pseudospin-$ \frac12$ Fermi surface has symmetry-protected point nodes along sixfold rotation axis. More specifically, we find that the low-energy gap structure$\Delta_{\bq}$ is given by
\be
\Delta_{\bq}  =  \frac{iA}{\sqrt{2}} (q^3_+-q^3_-) s_z +iB(q^2_+s_+-q^2_-s_-).  \label{eq:D6dirac2}
\ee
This shows that odd-parity $\ket{\bdelta_{D_6}}$ states can realize double Dirac points: low-energy gapless quasiparticles with quadratic dispersion in the $x$ and $y$ directions. 

The presence of mirror symmetry implies that the pseudospin-$ \frac32$ Fermi surface must have point nodes somewhere, even if they are not located along the sixfold rotation axis (where they might be expected). The location of these point nodes can be determined with the help of the mirror plane perpendicular to the sixfold axis. Indeed, the intersection of the three vertical mirror planes and $xy$ mirror plane defines six points on the Fermi surface, located along the $(010)$ and equivalent directions, which remain gapless. The dispersion of the gapless quasiparticles is linear,

As a third example of dihedral states, consider states with $D_4$ symmetry. Two examples are given by
\beq
\ket{\bdelta_{D_4}}_2 &=&  \ket{2,2} + \ket{2,-2}, \label{eq:D42}\\
\ket{\bdelta_{D_4}}_4 &=& \ket{4,2} + \ket{4,-2}. \label{eq:D44}
\eeq
The generators of the isotropy group are given by $\{e^{i\pi}C_{4z},C_{2x} \}$, and all pairing states with $D_4$ symmetry are time-reversal invariant. The odd-parity pairing states have a mirror symmetry $M_{(110)} \,:\, (x,y) \to (y,x)$ and its equivalent related by twofold rotation. This implies point nodes along the fourfold axis. The dispersion of the nodal quasiparticles is derived in the same way as before; we obtain the gap structure $\Delta_{\bq}$ for momenta $\pm \bK$ along the fourfold $z$-axis as
\be
\Delta_{\bq}  =  \frac{iA}{\sqrt{2}} (q^2_+-q^2_-) s_z +iB(q_-s_+-q_+s_-).  \label{eq:D6dirac3}
\ee
This shows that the gapless low-energy quasiparticles of odd-parity $\ket{\bdelta_{D_4}}$ pairing states have linear dispersion to lowest order and are yet another realization of Dirac superconductors. (Note that pseudospin-$\pm \frac12$ pairing also gives rise to linear dispersion.)

Finally, as an example of dihedral pairing states which break time-reversal symmetry, consider the states 
\beq
\ket{\bdelta_{D_3}}_3 &=& x_-\ket{3,3}  +2x_+ \ket{3,0}+x_-\ket{3,-3} , \label{eq:D33}\\
\ket{\bdelta_{D_3}}_4 &=& x_-\ket{4,3}  +2x_+ \ket{4,0}+x_-\ket{4,-3} , \label{eq:D33}\label{eq:D34}
\eeq
with $D_3$ symmetry. Here, $x_\pm \equiv \sqrt{1\pm x}$ and $x$ is a parameter which will depend on the details of the free energy. (These are therefore noninert states.) The generators of the isotropy group of both states are given by $\{C_{3z},e^{i\pi}C_{2x} \}$. Even though these states do not have a chirality, they break time-reversal symmetry and hence define spin-selective pairing states. Let us focus on the odd-parity realizations of these $D_3$ pairing states. We then notice that along the threefold axis, the gap  functions $\Delta_{\bq \pm } $ of pseudospin-$\frac32$ triplet pairing can have a constant part, i.e., $\Delta_{\bq \pm } \propto 1$, since the corresponding pseudospin-triplet pairing operators are invariant under threefold rotation. This implies a full pairing gap along the rotation axis. In contrast, for $\Delta_{\bq \pm } \propto q_\mp$ for pseudospin-$\frac12$ triplet pairing, giving rise to Majorana fermions with linear dispersion on the rotation axis.

\subsection{Application: cubic crystal anisotropy  \label{ssec:cubicsymmetry}}

Following the detailed exposition of pairing states with discrete symmetry, we conclude this section by demonstrating how the gap structure classification may be directly applied to systems with a normal state exhibiting crystal anisotropy. We focus the discussion on the cubic group, since one of the main motivations of this work are the half-Heusler materials. (We recall that the splitting of the isotropic channels in terms of cubic channels is listed in Table \ref{tab:cubicsymmetry}.)

As discussed in Sec. \ref{sec:pairingclass}, when crystal anisotropy effects reduce the spatial symmetry group of the (spin-orbit coupled) normal state to the crystal point group, pairing channels are labeled by representations of the crystal point group, and are necessarily finite dimensional. (Recall that the cubic representations have dimension one, two, or three.) In manner fully analogous to Secs. \ref{sec:pairingclass} and \ref{sec:GLtheory}, one may determine the set of stationary pairing states within each channel using symmetry arguments. The isotropy groups, which can be taken as a definition of distinct stationary pairing states, are necessarily discrete, since they must be subgroups of the normal state symmetry group. A complete list of cubic stationary states and their isotropy groups has been given by Volovik and Gorkov \cite{volovik85}.

For the purpose of deriving symmetry-enforced constraints on the gap structure of stationary pairing states, only the isotropy group is needed. The explicit form of the gap function is not required. This is important, since gap functions can be rather complicated in crystal systems due to the infinitely many symmetry-allowed terms within a representation (i.e., the available symmetry quantum numbers are greatly reduced in crystal point groups). In this regard, as far as the question of manifest (symmetry-enforced) gap structure properties is concerned, the question whether the normal state has full rotational symmetry or discrete crystal symmetry is secondary. What matters is the structure of the symmetry group of the pairing state; if it is discrete, as must be case with a cubic normal state, the theory of Sec. \ref{ssec:gapdiscrete} applies.

To make this more specific, consider the cubic normal state (spatial) symmetry group $O_h = O \times P$. Pairing states can be distinguished based on the parity eigenvalue, and for definiteness here we restrict to odd-parity pairing states. Then, there are five distinct pairing channels, labeled by the cubic representations $A_1$, $A_2$, $E$, $T_1$, and $T_2$ (see Sec. \ref{sec:pairingclass}; here we suppress the odd-parity designation). The $A_1$ and $A_2$ pairing channels are single-component channels, and therefore give rise to free energies with one unique stationary point. The symmetry of the pairing states then follows directly from the representations; the isotropy groups are generated by $\{C_{4z},C_{2,z+x} \}$ and $\{e^{i\pi}C_{4z},e^{i\pi}C_{2,z+x} \}$, respectively. These symmetry groups may be recognized as the groups of the two octahedral states of Sec. \ref{sssec:O} (see also Table \ref{tab:cubicsymmetry} in appendix \ref{app:cubic}), which implies that the gap structure is identical. 

The $E$ pairing channel is two-dimensional and the corresponding order parameter can be written as $\bdelta = (\Delta_{3z^2-r^2}, \Delta_{x^2-y^2})^\trans$ in the basis of Eq. \eqref{eq:scstate}. The GL functional for the two-component order parameter has three minima; two of them are given by $\bdelta = (1,0)^\trans$ and $\bdelta = (0,1)^\trans$. Both pairing states have dihedral symmetry group $D_4$, but in the former case it is generated by $\{C_{4z},C_{2x} \}$, whereas in the latter case $D_4$ is generated by $\{e^{i\pi}C_{4z},C_{2x} \}$. As a result, the state $\bdelta = (0,1)^\trans$ has the same isotropy group and gap structure as the $\ket{\bdelta_{D_4}}$ states of \ref{sssec:Dn} (see Table \ref{tab:cubicsymmetry}). In contrast, as discussed in Sec. \ref{ssec:gapdiscrete}, odd-parity pairing states with isotropy groups generated by pure rotations (i.e., no phase factors) are fully gapped due to the absence of constraints deriving from mirror symmetry. This directly applies to $\bdelta = (1,0)^\trans$.

As a final example, consider the three-component pairing channel $T_2$. A superconducting order parameter can be defined as  $\bdelta = (\Delta_{yz}, \Delta_{zx}, \Delta_{xy})^\trans$ [again in the basis of Eq. \eqref{eq:scstate}]. In total, four distinct pairing states can arise in the $T_2$ channel. For the purpose of illustration, here we just consider two: the time-reversal invariant state $\bdelta = (1,1,1)^\trans$ and chiral state $\bdelta = (1,i,0)^\trans$. The former has dihedral isotropy group $D_3$ generated by $\{C_{3,x+y+z}, C_{2,x-y} \}$, which per Sec. \ref{ssec:gapdiscrete} implies a full (topological) pairing gap. The time-reversal odd pairing state is left invariant under the group generated by $e^{i\pi/2}C_{4z}$, implying that $\bdelta = (1,i,0)^\trans$ is chiral and has (axial) angular momentum $+1$ along the threefold axis. The gap structure may then be obtained by applying the arguments of Sec. \ref{sssec:triplet} to discrete $n$-fold rotations. 

To summarize these considerations, even in cases where the normal state has discrete crystal symmetry, the gap structure classification developed in this section can be directly applied, since pairing states with discrete symmetry are naturally included. We have demonstrated this explicitly using the example of the cubic group, but the conclusion holds for any other crystal point group.

\section{Discussion and Conclusion \label{sec:conclusion}}

In this work we have presented a comprehensive topological gap structure classification of $j=\frac32$ pairing states, obtained through a systematic analysis of the constraints enforced by symmetry. Our analysis of multicomponent pairing states demonstrates that in strongly spin-orbit coupled systems with higher total angular momentum pairing, and in particular in systems with high-spin pairing, topological pairing states form a significant subset of the class of possible superconducting ground states. Four broad classes of topological pairing states should be distinguished: fully gapped time-reversal invariant topological superconductors, nodal Dirac superconductors, nodal superconductors hosting Majorana fermions, and superconductors with $\mathbb{Z}_2$ protected Bogoliubov Fermi surfaces. 

Within each class, further distinctions can be made. For instance, fully gapped topological superconductors can be either isotropic or nematic. Nematic superconductors spontaneously break rotation symmetry and have an anisotropic pairing gap \cite{fu14}. The latter provides a useful experimental diagnostic, as it does not require phase sensitive probes. Within the class of superconductors with bulk nodal gapless excitations, pairing states can be distinguished based on the Berry monopole charge of the point nodes, where the monopole charge is directly related to the dispersion of the low-energy quasiparticles. For instance, Dirac or Majorana quasiparticles with linear dispersion are different from quasiparticles with quadratic dispersion, and define distinct superconducting states.   

Our work shows that Majorana fermions generically occur in spin-orbit coupled $j=\frac32$ superconductors with multicomponent odd-parity pairing which spontaneously breaks time-reversal symmetry. As discussed in Sec. \ref{ssec:spinselective}, spin-orbit coupled superconductors with broken time-reversal symmetry are generically spin-selective, allowing for an effective pairing-induced pseudospin splitting. This splitting is responsible for the lifting of pseudospin degeneracies of nodal points on rotation axes, thereby giving rise to non-degenerate point nodes. As far as multicomponent even-parity superconductors are concerned, the spin-selectiveness of the pairing implies that gap structures generically feature the $\mathbb{Z}_2$ surface degeneracies \cite{agterberg17}. 

The gap structure classification we establish in this work provides a useful framework for interpreting ongoing and future experiments which target bulk properties of superconductors. In particular, thermodynamic probes such as specific heat, penetration depth, or NMR spin relaxation time measurements are sensitive to the nature of low-energy excitations \cite{matsuda06}. The low-temperature behavior of these quantities directly reflects the density of low-energy quasiparticle states. More specifically, whereas fully gapped superconductors exhibit exponentially activated temperature dependence, nodal superconductors exhibit a power-law dependence at temperatures $T \ll T_c$. The power-law exponent is directly related to the low-energy quasiparticle density of states, and therefore allows to distinguish nodes with different codimension. Notably, however, the density of states of point nodes depends on the Berry monopole charge, which can give rise to low-temperature behavior expected for nodes of different codimension. As a notable example, point nodes with quadratic dispersion can masquerade as line nodes. Our classification is therefore directly useful for the purpose of assigning candidate pairing states to experimentally observed behavior. It is also worth pointing out that for odd-parity time-reversal symmetry breaking pairing states which host non-degenerate Majorana fermions, a further experimental signature is NMR spin relaxation time anisotropy \cite{kozii16}. 

The symmetry properties of pairing states are fundamental to our classification of gap structures. These symmetries properties are uniquely encoded in the subsidiary order parameters associated with the superconducting state, which take the form magnetic multipole moments (e.g., dipole moment or chirality; quadrupole moment). Therefore, information on the nature of the superconducting state becomes accessible by probing the structure of the multipole moments. Time-reversal symmetry and rotation symmetry breaking, for instance, can be determined by polar Kerr effect measurements and thermal conductivity or specific heat measurements as function of magnetic field direction, respectively.  

The defining physical manifestation of bulk topology are the gapless excitations on the boundary of the material. Topological superconductors with a full pairing gap host two-dimensional gapless Majorana fermions on their surfaces. The existence of these surface Majorana fermions does not depend on surface termination. However, in the case of nematic superconductors, the precise form of the surface quasiparticle dispersion is expected to be anisotropic and sensitive to surface termination with respect to the nematic axis. Bulk nodal superconductors are characterized by gapless Majorana arc surface states, which connect the projections onto the surface Brillouin zone of bulk nodes with opposite monopole charge \cite{meng12}. As a result, their structure is inherently surface termination dependent. In Dirac superconductors, which are time-reversal invariant and possess a mirror symmetry, these Majorana arcs must come in pairs: Majorana-Kramers pairs \cite{zhang13,yang14}. 

The gapless surface excitations can be probed using tunneling microscopy experiments, which couple to the surface density of states. An interesting direction for future work is to study the surface tunneling spectra for different pairing states and different surface terminations. 

We conclude this paper by pointing out two important implications of our work. First, since the formalism of our gap structure classification includes pairing states with discrete symmetry, it encompasses the pairing ground states which can arise when crystal anisotropy effects are accounted for. As a result, insofar as the question of quasiparticle gap structure is concerned---the primary interest of this work---the application of our classification is not limited to superconductors with full rotational symmetry. Furthermore, despite our focus on $j=\frac32$ pairing in the Luttinger model, our topological gap structure classification is directly relevant to higher angular momentum pairing in a more general setting, in particular other systems with strong spin orbit coupling. Symmetry arguments are the work horse of our approach and we therefore expect our analysis of multicomponent topological pairing states to find broad application.

{\it ---Note added.} After finalization of this manuscript we became aware of a preprint which also considers topological superconducting states in the Luttinger models \cite{roy17}.

\begin{acknowledgements}
J.V. and L.F. wish to thank Vladyslav Kozii for previous collaborations related to this work. L.S. and J.R. were supported by the Gordon and Betty Moore Foundation through scholarships of the EPiQS initiative under grant no.\ GBMF4303. L.S. acknowledges the hospitality of the KITP and NSF grant PHY-1125915. P.A.L. was supported by the DOE under grant no.\ FG02-03ER46076. L.F. and J.V. acknowledge funding by the DOE Office of Basic Energy Sciences, Division of Materials Sciences and Engineering under Award No. DE-SC0010526. 
\end{acknowledgements}

\appendix

\section{Spin multipole matrices of $j=\frac32$ fermions \label{app:spinmat}}

The spin matrices $\bS=  (S_x, S_y,S_z)$ of the $j=\frac{3}{2}$ multiplet are given by
\be
S_z  =  \begin{pmatrix}  \frac{3}{2} & 0 & 0 &0 \\ 0& \frac{1}{2} &0 &0 \\ 0&0&  -\frac{1}{2} &0  \\ 0 & 0 &0 &-\frac{3}{2}  \end{pmatrix} , \;  S_+ = S_-^\dagger = \begin{pmatrix}  0 & \sqrt{3} & 0 &0 \\ 0& 0 &2 &0 \\ 0&0&  0 &\sqrt{3}  \\ 0 & 0 &0 &0  \end{pmatrix} \label{eq:Smat}
\ee
where $S_\pm = S_x \pm i S_y$. The spin multipole matrices $ \mathsf{ S}_{SM}$ introduced in Eq. \eqref{eq:pairSMS}, where $S$ is the total spin of two $j=\frac{3}{2}$ fermions, and $M$ is their total axial spin angular momentum, can be obtained as follows. First, notice that the matrices $\mathsf{ S}_{1M}$ are proportional to the spin matrices $S_{\pm}$ and $S_z$ of Eq. \eqref{eq:Smat}. Specifically, one has
\be
 \mathsf{ S}_{1\pm 1} = \mp \frac{1}{\sqrt{10}} S_{\pm}, \quad  \mathsf{ S}_{10} =  \frac{1}{\sqrt{5}} S_z. 
\ee
The higher order multipole matrices $ \mathsf{ S}_{SM}$, where $S=2,3$, can be obtained applying the recursive formula
\be
[S_-,  \mathsf{ S}_{SM}] = \sqrt{S(S+1)-M(M-1)} \mathsf{ S}_{S,M-1},
\ee
to the highest weight matrix with $M=S$. For each $S$, the highest weight matrix is obtained by setting $\mathsf{ S}_{SS} \propto \mathsf{ S}^S_{11}$ and requiring that the normalization of the matrices $\mathsf{ S}_{SM}$ is such that the sum rules
\be
c^\dagger_{\bk\alpha}c^\dagger_{-\bk\beta} = \sum_{S,M} \Big\langle\frac32\frac32, SM \Big| \frac32\frac32,\alpha\beta\Big\rangle \Pi^\dagger_{SM}(\bk),
\ee
are satisfied, where $ \bracket{\frac32\frac32,SM}{\frac32\frac32,\alpha\beta} $ are the Clebsch-Gordan coefficients. This is satisfied by the normalization condition (no sum over $M$)
\be
\tr{\spinS{S}{M}\spinS{S}{M}^\dagger}=1.
\ee

The matrices $\spinS{S}{M}$ encode the spin multipole structure of the Cooper pair. Since the total spin of the Cooper pair can be $S=1,2,3$ (apart from $S=0$), Cooper pairs can have spin dipole, quadrupole and octupole moments.
To highlight the interpretation of spin multipole moments, we take $S=2$ as an and construct the multipole components contained in the set $\mathsf{S}_{2M}$ explicitly. The spin quadruple matrices are defined by a rank-2 symmetric traceless tensor $Q_{ab}$, where $a,b\in \{x,y,z\}$, given by
\begin{gather}
Q_{ab} = \frac{1}{2}(S_aS_b+ S_aS_b) -\frac{5}{4}\delta_{ab} . \label{eq:quadrupole}
\end{gather}
Symmetric and traceless tensors such as $Q_{ab}$ have five independent components, which precisely matches the number of $S=2$ matrices $\mathsf{S}_{2M}$. The explicit linear correspondence between the five components of $Q_{ab}$ and $\mathsf{S}_{2M}$ is presented in Table \ref{tab:quadrupolar}. Note that $Q^\dagger_{ab} =Q_{ba}$ and therefore the quadruple components are real. 

\begin{table}[t]
\centering
\begin{ruledtabular}
\begin{tabular}{ccc}
Components of $S_{2M}$  & Components of $Q_{ab} $ & Cubic  \\ [4pt]
\hline  
$\mathsf{S}_{22}+\mathsf{S}_{2-2}$ &  $Q_{xx} - Q_{yy}$ & $E_{g,1}$  \\
$\sqrt{2}\mathsf{S}_{20}$ & $\frac{1}{\sqrt{3}} (2 Q_{zz} - Q_{xx} - Q_{yy} ) $ & $E_{g,2}$ \\
$\mathsf{S}_{2-1}-\mathsf{S}_{21}$ &  $Q_{xz}+Q_{zx} $ & $T_{2g,1}$ \\
$-i(\mathsf{S}_{21}+\mathsf{S}_{2-1})$& $ Q_{yz}  +Q_{zy} $ & $ T_{2g,2}$ \\
$-i(\mathsf{S}_{22}-\mathsf{S}_{2-2})$&  $Q_{xy} +Q_{yx} $ & $ T_{2g,3}$
\end{tabular}
\end{ruledtabular}
 \caption{Correspondence between the spin matrices $S_{2M}$ and the components of the rank-2 tensor $Q_{ab} $, demonstrating that $S_{2M}$ transform as a rank-2 tensor. $S_{3M}$ (not shown) transform as a rank-3 tensor.
 }
\label{tab:quadrupolar}
\end{table}

\section{General angular momentum multipole matrices and subsidiary orders \label{app:multipole}}

The notion of angular momentum multipole matrices is also at the heart of the definition of the subsidiary orders $I_{KN}$ introduced in Sec. \ref{ssec:GLgeneral}. Recall that $I_{KN}$ are defined as 
\be
I_{KN} =\bdelta^\dagger   \mathcal{I}_{KN}  \bdelta = \sum_{MM'}(\mathcal{I}_{KN})_{MM'}\Delta^*_M \Delta_{M'},
\ee
where we have chosen the chiral basis for the matrices $\mathcal{I}_{KN}$; $M,M'$ are magnetic quantum numbers of a superconductor with total angular momentum $J$. Note that $J$ is always integer (and not half-odd integer). The matrices $\mathcal{I}_{KN}$ have dimensions $(2J+1)\times(2J+1)$, and can be constructed in the same way as $\mathsf{S}_{SM_S}$. The dipole matrices $\mathcal{I}_{1,N=1,0,-1}$ are proportional to linear combinations of the three spin matrices $\mathcal{I}_{x,y,z}$, where $(\mathcal{I}_{z})_{MM'}=M\delta_{MM'}$ and $\mathcal{I}_{1,\pm1} =\mp \mathcal{I}_{\pm}/\sqrt{2} = \mp(\mathcal{I}_{x}\pm i \mathcal{I}_{y})/\sqrt{2}$.

Higher order multipole matrices are obtained by first constructing the highest weight state $\mathcal{I}_{KK}\propto (\mathcal{I}_{11})^K$, normalizing, and then using 
\begin{gather}
[ \mathcal{I}_-,  \mathcal{I}_{KN}] = \sqrt{K(K+1)-N(N-1)} \mathcal{I}_{K,N-1}, \label{eq:mmultipole2}
\end{gather}
Recall that the highest multipole possible is $K=2J$, implying that for a total angular momentum $J$ superconductor $2J$ distinct subsidiary order parameters can be defined.

\section{Invariants of the Ginzburg-Landau free energy functional \label{app:GLinvariant}}

In this appendix we show that the sum over $K$ in Eq. \eqref{eq:FJ} of the main text contains $J$ terms, i.e., $K=1,\ldots,J$. For this purpose it is convenient to choose the chiral basis $\Delta_{M}$ for the superconducting order parameters, see Eq. \eqref{eq:spinstate}. 

In its most general form, the fourth order contribution to the GL free energy density can be written as a quartic interaction of the order parameter fields. Such interaction can be written as
\be
f^{(4)}_{J} = \sum_{MNPQ} \hat V_{MNPQ}\Delta^*_M\Delta^*_N\Delta_P\Delta_Q.  \label{eqapp:GLint}
\ee
(Here $M,N,P,Q$ are all magnetic angular momentum indices.)  In this form, we may interpret $f^{(4)}_{J}$ as a pair scattering interaction: a $(PQ)$ pair is scattered to a $(MN)$ pair with scattering vertex $\hat V_{MNPQ}$. Borrowing knowledge from the theory of spinor Bose-Einstein condensates \cite{ho98}, the interaction $\hat V$ can be decomposed into channels of total angular momentum $\tilde K$, where---and this is important---$\tilde K$ refers to the total angular momentum of a pair $\Delta_M\Delta_N$ ($K$ always refers to the total angular momentum of a gauge-invariant bilinear $\Delta^*_M\Delta_N$), expressed as
\be
\hat V = \sum_{{\tilde K}}  \hat V_{{\tilde K}} \mathcal{P}_{{\tilde K}}. \label{eqapp:GLprojector}
\ee
Here $\hat V_{{\tilde K}}$ are real interaction parameters and $\mathcal{P}_{{\tilde K}} = \sum_{M_{{\tilde K}}}\ket{{\tilde K},M_{{\tilde K}}}\bra{{\tilde K},M_{{\tilde K}}}$ projects the pairs onto a total angular momentum ${\tilde K}$ state, such that the matrix elements $\hat V_{MNPQ}$ are given by
\be
\hat V_{MNPQ} = \sum_{{\tilde K} =0,2,\ldots}  \hat V_{\tilde K}  \bra{MN} \mathcal{P}_{\tilde K}  \ket{PQ}, 
\ee
Since the $\Delta_{M}$ are complex commuting fields, $\tilde K$ must be even. The maximal value of $\tilde K$ equals $2J$, yielding a total of $J+1$ distinct terms in \eqref{eqapp:GLprojector}. This establishes that $f^{(4)}_{J}$ is parametrized by $J+1$ independent interaction coefficients $\hat V_{\tilde K} $. 

This matches the number of interaction coefficients of Eq. \eqref{eq:FJ} given by $(u,v_K)$, but it does not, however, prove that $f^{(4)}_{J}$ takes the exact form of Eq. \eqref{eq:FJ}, with subsidiary order parameters $I_{KM_K}$ given by Eq. \eqref{eq:subsorder}. To show this, we first note the identity
\be
\sum_{\tilde K=0,2,}^{2J} [\tilde K(\tilde K+1)-2J(J+1) ]^n \mathcal{P}_{\tilde K}  = (2 \boldsymbol{\mathcal{I}}^\dagger \cdot  \boldsymbol{\mathcal{I}})^n,
\ee
for each $n$, where $\boldsymbol{\mathcal{I}} = (\mathcal{I}_{11},\mathcal{I}_{10},\mathcal{I}_{1-1})$, or, equivalently, $\boldsymbol{\mathcal{I}} = (\mathcal{I}_{x},\mathcal{I}_{y},\mathcal{I}_{z})$. In the latter case the matrices satisfy $\boldsymbol{\mathcal{I}}^\dagger=\boldsymbol{\mathcal{I}}$ and the choices of basis are related by
\be
\mathcal{I}_{z}= \mathcal{I}_{10}, \quad \mathcal{I}_{1\pm 1}= \mp(\mathcal{I}_{x}\pm i\mathcal{I}_{y})/\sqrt{2}.
\ee
Together with the identity
\be
\sum_{{\tilde K} =0,2,\ldots}^{2J} \mathcal{P}_{\tilde K}  = 1,
\ee
we now have $J+1$ equations relating $\mathcal{P}_{\tilde K} $, with $\tilde K=0,2,\ldots,2J$, to $(2 \boldsymbol{\mathcal{I}}^\dagger \cdot  \boldsymbol{\mathcal{I}})^n$, with $n=0,1,\ldots,J$. [Note that $(2 \boldsymbol{\mathcal{I}}^\dagger \cdot  \boldsymbol{\mathcal{I}})^0=1$.] We can thus write the interaction as
\be
\hat V = v_0 1 + \sum_{n=1}^J v_n (2 \boldsymbol{\mathcal{I}}^\dagger \cdot  \boldsymbol{\mathcal{I}})^n. \label{eqapp:GLint2}
\ee
This proves Eq. \eqref{eq:FJ}, since $v_0=u$ and each term $(\boldsymbol{\mathcal{I}}^\dagger \cdot  \boldsymbol{\mathcal{I}})^n$ can always be expressed as a linear combination of terms of the form $\sum_{M_K} \mathcal{I}^\dagger_{KM_K}\mathcal{I}_{KM_K}$, where $K$ can take values between $0$ and $n$. Explicitly, one has
\be
(\boldsymbol{\mathcal{I}}^\dagger \cdot  \boldsymbol{\mathcal{I}})^n = \sum_{K=0}^n c_K \sum_{M_K} \mathcal{I}^\dagger_{KM_K}\mathcal{I}_{KM_K}.
\ee


\section{Band basis operators \label{app:projection}}

The operators which create and annihilate states in the band basis are defined as $f^{(\dagger)}_{\bk\mu}$ (valence band) and  $d^{(\dagger)}_{\bk\mu}$ (conduction band). Here, $\mu$ labels the pseudospin degree of freedom of the two bands, $\pm \frac32$ and $\pm \frac12$, denoted as $\mu=\up,\down$. We require that the basis for this pseudopsin is chosen such that $\ket{\bk, \mu}$ transform under $\Theta$ and $P$ as an a usual spin. This implies
\beq
P  \ket{\bk, \mu}  &=&   \ket{-\bk, \mu}   \label{appeq:Pproject} \\
\Theta  \ket{\bk, \mu}  &=& \epsilon_{\mu\nu}  \ket{-\bk, \nu} . \label{appeq:TRSproject}
\eeq
Taking the valence band as an example, the matrix which relates the operators $f_\bk$ and $c_\bk$ is defined as $V_\bk$ and can be explicitly represented as 
\be
V_\bk = \begin{pmatrix}  {\bf v}_{1} & {\bf v}_{2}  \end{pmatrix} \label{appeq:umatrix}
\ee
where ${\bf v}_{1,2} $ are the vectors of the $\ket{\bk, \mu} $ states in the basis of $c^\dagger_{\bk\alpha} \ket{0} $. Note that this makes $V_\bk$ a $4\times 2$ matrix. The relation between $f_\bk$ and $c_\bk$ then reads as
\be
f^\dagger_{\bk\mu} = c^\dagger_{\bk\alpha}(V_\bk)_{\alpha\mu}, \quad \hat{\mathcal{P}}^vc^\dagger_{\bk\alpha}\hat{\mathcal{P}}^v=f^\dagger_{\bk\mu}(V^\dagger_\bk)_{\mu\alpha}. \label{appeq:ctof}
\ee
Here, $\hat{\mathcal{P}}^v$ is the projection operator onto the valence band states, i.e., it projects out operators of the conduction band. Now, the symmetry requirements of Eqs. \eqref{appeq:Pproject} and \eqref{appeq:TRSproject} can be formulated in terms of the eigenvector matrix $U_\bk$. Using that $c_\bk$ transforms under time-reversal as $\Theta c_{\bk \alpha} \Theta^{-1} = \Tm_{\alpha\beta} c_{\bk \beta}$, we find Eq. \eqref{appeq:TRSproject} implies
\be
\Tm^\trans V^*_\bk \epsilon = V_{-\bk},
\ee
where $\epsilon \equiv i\sigma_y$. Note that $\epsilon^*=\epsilon$ and $\epsilon^\trans=-\epsilon$. The requirement of inversion symmetry is simply $V_\bk=V_{-\bk}$, which is trivially satisfied since the Hamiltonian is even under inversion. 

Similarly, the matrix of conduction band eigenvectors is defined as $W_{\bk}$, and we have
\be
d^\dagger_{\bk} = c^\dagger_{\bk}W_\bk, \quad d_{\bk} = W^\dagger_\bk c_{\bk}. 
 \label{eq:ctod}
\ee
The quasiparticle operators $c_{\bk}$ and $c^\dagger_{\bk}$ can then be expressed in terms of $f_{\bk}$ and $d_{\bk}$ as
\be
c_{\bk}  =  V_\bk f_{\bk} + W_\bk d_{\bk}, \quad  c^\dagger_{\bk}  =  f^\dagger_{\bk} V^\dagger_\bk + d^\dagger_{\bk}W^\dagger_\bk  .
 \label{appeq:ctofd}
\ee

A set of basis vectors ${\bf v}_{1,2} $ and ${\bf w}_{1,2} $ may be found by diagonalizing $(\bk \cdot \bS)^2$ and choosing the eigenvectors such that the requirements of Eqs. \eqref{appeq:Pproject} and \eqref{appeq:TRSproject} are satisfied. The basis vectors can be specified in terms of the $L=1$ spherical harmonics $Y_{1M}$ as
\be
 \begin{pmatrix}  {\bf v}_{1} & {\bf v}_{2}  \end{pmatrix}  =  |Y_{11}|\begin{pmatrix} -Y_{10}/Y_{11} &  \frac{1}{\sqrt{2}}Y^*_{11}/Y_{11}\\ \sqrt{\frac{3}{2}} & 0 \\0 & \sqrt{\frac{3}{2}}   \\ \frac{1}{\sqrt{2}}Y_{11}/Y^*_{11}& Y_{10}/Y^*_{11} \end{pmatrix} . \label{appeq:basisvalence}
\ee
Observe that even though the $L=1$ spherical harmonics are odd under inversion, these states satisfy ${\bf v}_{1,2}(\bk) = {\bf v}_{1,2} (-\bk)$, and therefore the matrix $V_\bk$ of Eq. \eqref{appeq:umatrix} constructed from these states trivially satisfies $V_\bk=V_{-\bk}$. In the same way, for ${\bf w}_{1,2} $ one has
\be
 \begin{pmatrix}  {\bf w}_{1} & {\bf w}_{2}  \end{pmatrix}  = \frac{1}{\sqrt{N}} \begin{pmatrix} -\sqrt{3} Y_{10}Y^*_{11} & - \sqrt{\frac{3}{2}}(Y^*_{11})^2 \\ -N & 0 \\0 &N  \\ \sqrt{\frac{3}{2}}Y^2_{11} & -\sqrt{3} Y_{10}Y_{11}  \end{pmatrix}, \label{appeq:basisconduction}
\ee
where $N= 2Y^2_{10}+|Y_{11}|^2$.

To consider the action of spatial symmetries on the pseudospin operators $f_{\bk}$ and $d_{\bk}$ we denote an element of $O(3)$ as $R$. The spin $j=\frac32$ quasiparticle operators $c^\dagger_{\bk}$ transform under $R$ as
\be
\hat{R} c^\dagger_\bk \hat{R}^\dagger =  c^\dagger_{R \bk}U_{R} , 
\ee
where $U_{R}$ is the $j=\frac32$ matrix representation of $R$. We define the matrix representation of $R$ on the pseudospin degree of freedom $f^\dagger_{\bk\mu}$ as $O_{R}(\bk)$, which in general will depend on $\bk$. Then, we find that $O_{ R}(\bk)$ is related to $U_R$ as
\be
V_{R\bk}O_{R}(\bk) = U_R V_{\bk} ,
\ee
from which we obtain $O_{R}(\bk) $ as
\be
O_{R}(\bk) = V^\dagger_{R\bk}  U_R V_{\bk} .
\ee
Clearly, a similar relation holds for $W_\bk$.

\section{Low-energy quasiparticle gap structure: Explicit projection \label{app:qpproject}}

In this appendix we derive general expressions for the projected pairing. Specifically, given a pairing potential $\Delta_\bk$ in  Eq. \eqref{eq:HBdGk} we project onto the low-energy Fermi surface degrees of freedom at special momenta $\pm \bK$. We start by decomposing the quasiparticle operators $c_\bk$ at $\pm \bK$ in terms of valence band and conduction band operators $f_\bk$ and $d_\bk$; we find
\be
c_{\pm \bK}  =  V f_{\pm \bK} + W d_{\pm \bK}, \quad  c^\dagger_{\pm \bK}  =  f^\dagger_{\pm \bK} V^\dagger + d^\dagger_{\pm \bK}W^\dagger ,
 \label{eq:ctofdq}
\ee
where $V \equiv V_\bK = V_{-\bK}$ and $W \equiv W_\bK = W_{-\bK}$ are the matrices of eigenvectors, see Eq. \eqref{eq:ctofd}. With the help of these relations we expand the pairing Hamiltonian in the vicinity of $\pm \bK$ as
\begin{multline}
\mathcal{H} \simeq \frac12 \sum_{\bq}  \Psi^\dagger_{\bq } \mathcal{H}^{vv}_\bq \Psi_{\bq  }+\frac12\sum_{\bq}  \Phi^\dagger_{\bq } \mathcal{H}^{cc}_\bq \Phi_{\bq  } \\
+\frac12 \sum_{\bq}  \Psi^\dagger_{\bq } \mathcal{H}^{vc}_\bq \Phi_{\bq  }+\frac12\sum_{\bq}  \Phi^\dagger_{\bq } \mathcal{H}^{cv}_\bq \Psi_{\bq  } , \label{eq:LEHproject}
\end{multline}
where $\Psi_\bq$ and $\Phi_\bq$ were defined in Eqs. \eqref{eq:Psiq} and \eqref{eq:Gammaq}.  The Hamiltonian components $ \mathcal{H}^{vv}_\bq$ and $\mathcal{H}^{vc}_\bq$ (\emph{v} and \emph{c} label the valence and conductions bands, respectively) are given by the matrix expressions
\begin{widetext}
\be
 (\mathcal{H}^{vv}_\bq)_{\mu\nu}   =  \begin{pmatrix} \varepsilon^{v}_{\bq} \delta_{\mu\nu}  & 0  & 0   &  (V^\dagger \Delta_\bq \Tm V^*)_{\mu\nu} \\ 0 & \varepsilon^{v}_{-\bq} \delta_{\mu\nu}  & \pm (V^\dagger \Delta_{-\bq} \Tm V^*)_{\mu\nu} & 0  \\ 0  &  \pm (V^\trans \Tm^\trans\Delta^\dagger_{-\bq} V)_{\mu\nu}& - \varepsilon^{v}_{-\bq} \delta_{\mu\nu} & 0  \\  (V^\trans \Tm^\trans\Delta^\dagger_\bq V)_{\mu\nu} & 0&0 &  -\varepsilon^{v}_{\bq} \delta_{\mu\nu}   \end{pmatrix},\label{eq:Hvv}
\ee
where we have defined $\Delta_{\pm \bq} \equiv \Delta_{\bK \pm \bq}$ as in the main text, and
\be
 (\mathcal{H}^{vc}_\bq)_{\mu\nu}   =  \begin{pmatrix} 0 & 0  & 0   &  (V^\dagger \Delta_\bq \Tm W^*)_{\mu\nu} \\ 0 & 0 & \pm (V^\dagger \Delta_{-\bq} \Tm W^*)_{\mu\nu} & 0  \\ 0  &  \pm (V^\trans \Tm^\trans\Delta^\dagger_{-\bq} W)_{\mu\nu}& 0 & 0  \\  (V^\trans \Tm^\trans\Delta^\dagger_\bq W)_{\mu\nu} & 0&0 & 0   \end{pmatrix}.\label{eq:Hvc}
\ee
\end{widetext}
The Hamiltonian blocks $ \mathcal{H}^{cc}_\bq$ and $\mathcal{H}^{cv}_\bq$ are simply obtained from Eqs. \eqref{eq:Hvv} and \eqref{eq:Hvc}, respectively, by substituting $V \leftrightarrow W$. In these expressions $\pm$ applies to even-parity ($+$) and odd-parity ($-$) pairing states. 

Since we are assuming a valence band Fermi surface, the conduction band defines a high-energy manifold. To project the pairing onto the valence band subspace close to $\pm \bK$ we apply perturbation theory. An effective Hamiltonian $\mathcal{H}^{vv}_{\text{eff}} (\bq)$ for the valence band subspace is given by an expression similar to Eq. \eqref{eq:Hvveff} as 
\be
\mathcal{H}^{vv}_{\text{eff}} (\bq) \simeq \mathcal{H}^{vv}_\bq -  \mathcal{H}^{vc}_\bq (\mathcal{H}^{cc}_\bq)^{-1}\mathcal{H}^{cv}_\bq, \label{eq:Hperturb}
\ee
and we can expand $(\mathcal{H}^{cc}_\bq)^{-1}$ in powers of $(\varepsilon^{c}_{\bq=0})^{-1}$. Here, $\varepsilon^{c}_{\bq=0} = \varepsilon^{c}_{\bK}$ is the conduction band energy at the Fermi momentum $\bK$. The structure of Eq. \eqref{eq:Hvv} shows that $\mathcal{H}^{cc}_\bq$ is the sum of the normal state part and the pairing part, and can be expressed as
\be
\mathcal{H}^{cc}_\bq = \varepsilon^{c}_{\bK} \tau_z +\Delta X_\bq,
\ee
where we have neglected the $\bq$-dependence of the normal state contribution. The matrix $X_\bq$ describes the pairing part and $\Delta$ is the overall amplitude of the superconducting order parameter, which we may take to be real. With this we may expand $(\mathcal{H}^{cc}_\bq)^{-1}$ as
\be
(\mathcal{H}^{cc}_\bq)^{-1} = \frac{1}{\varepsilon^{c}_{\bK}} \left[ \tau_z - \frac{\Delta}{\varepsilon^{c}_{\bK}}  X_\bq  + \mathcal{O}\left(\frac{\Delta^2}{(\varepsilon^{c}_{\bK})^2}\right) \right].
\ee
Note that the expansion parameter $\Delta/\varepsilon^{c}_{\bK}$ is typically small. Substituting this expansion into Eq. \eqref{eq:Hperturb} we obtain
\begin{widetext}
\begin{multline}
-  \mathcal{H}^{vc}_\bq (\mathcal{H}^{cc}_\bq)^{-1}\mathcal{H}^{cv}_\bq= \frac{1}{\varepsilon^{c}_{\bK}} \begin{pmatrix} V^\dagger \Delta_\bq \mathcal{P}^c_\bK \Delta^\dagger_\bq V  & 0  & 0   & 0 \\ 0 & V^\dagger \Delta_{-\bq} \mathcal{P}^c_\bK \Delta^\dagger_{-\bq} V & & 0  \\ 0  &0  & -\epsilon^TV^\dagger \Delta^\dagger_{-\bq} \mathcal{P}^c_\bK \Delta_{-\bq} V\epsilon& 0  \\ 0 & 0&0 & - \epsilon^TV^\dagger \Delta^\dagger_{\bq} \mathcal{P}^c_\bK \Delta_{\bq} V\epsilon  \end{pmatrix}  +
\\
\frac{1}{(\varepsilon^{c}_{\bK})^2}  \begin{pmatrix}  0& 0  & 0   & V^\dagger \Delta_\bq \mathcal{P}^c_\bK \Delta^\dagger_\bq  \mathcal{P}^c_\bK \Delta_\bq V\epsilon  \\ 0 &0 & \pm V^\dagger \Delta_{-\bq} \mathcal{P}^c_\bK \Delta^\dagger_{-\bq}  \mathcal{P}^c_\bK \Delta_{-\bq} V\epsilon & 0  \\ 0  &\pm \epsilon^T V^\dagger \Delta_{-\bq} \mathcal{P}^c_\bK \Delta^\dagger_{-\bq}  \mathcal{P}^c_\bK \Delta^\dagger_{-\bq} V  &0 & 0  \\ \epsilon^T V^\dagger \Delta_\bq \mathcal{P}^c_\bK \Delta^\dagger_\bq  \mathcal{P}^c_\bK \Delta^\dagger_\bq V  & 0&0 & 0  \end{pmatrix}.\label{eq:Hperturbexpand}
\end{multline}
\end{widetext}
Here, $ \mathcal{P}^c_\bK \equiv WW^\dagger$ is the matrix projector onto the conduction band states at $\bK$ (and hence $-\bK$). In this expression, the first term can be recognized as a particle-hole term and is responsible for the effective Zeeman-type pseudospin splitting, see Eq. \eqref{eq:Fzeeman}. To obtain $\delta$ one sets $\bq=0$ in the this first term. The first term may also contain a renormalization of the single-particle energies. 

The second term describes a contribution to the pairing of valence band due to coupling to the conduction band. As a result, it is smaller by one order of $\Delta/\varepsilon^{c}_{\bK}$ and not expected to be of significance.

\section{Connection to cubic symmetry  \label{app:cubic}}

\begin{table*}[t]
\centering
\begin{ruledtabular}
\begin{tabular}{ccllc}
$J$  & Cubic symmetry  &   Gap functions ($\ket{J,M}$)  & Labels Sec. \ref{ssec:gapdiscrete} & Expr. \\ 
\hline
$1$ & $T_{1g,u}$ &    $\begin{aligned} \ket{x} & =  \tfrac{1}{\sqrt{2}}(\ket{1,-1}-\ket{1,1})\\  \ket{y} & =  \tfrac{1}{i\sqrt{2}}(\ket{1,-1}+\ket{1,1})\\  \ket{z} & =  \ket{1,0} \end{aligned}$  & & \\[22pt]
 \hline
$2$ & $E_{g,u} $  & $\begin{aligned} & \ket{ 3z^2-r^2}  =   \ket{2,0}   \\   & \ket{x^2-y^2}  =  \tfrac{1}{\sqrt{2}}(\ket{2,2}+\ket{2,-2})  \end{aligned}$ & $\begin{aligned} & \ket{\bdelta_{D_4}} = \ket{x^2-y^2} \\ &\ket{\bdelta_{T}} = \ket{x^2-y^2}+i\ket{ 3z^2-r^2}  \end{aligned}$ & $\begin{aligned} & \text{Eq. \eqref{eq:D42}} \\ & \text{Eq. \eqref{eq:tetra2}}  \end{aligned}$\\[15pt]
 & $ T_{2g,u}$ &  $\begin{aligned} \ket{yz} & =  \tfrac{1}{i\sqrt{2}}(\ket{2,-1}+\ket{2,1})\\  \ket{zx} & =  \tfrac{1}{\sqrt{2}}(\ket{2,-1}-\ket{2,1})\\  \ket{xy} & =  \tfrac{1}{i\sqrt{2}}(\ket{2,2}-\ket{2,-2}) \end{aligned}$ & & \\[22pt]
 \hline
$3$ & $A_{2g,u} $ & $ \ket{xyz} =  \tfrac{1}{i\sqrt{2}}(\ket{3,2}-\ket{3,-2})$ &  $\ket{\bdelta_{O}}_3$ &  Eq. \eqref{eq:octa2}\\[5pt] 
 & $T_{1g,u} $ & $\begin{aligned} \ket{x^3} & = \tfrac{\sqrt{5}}{4}(-\ket{3,3}+\ket{3,-3})+\tfrac{\sqrt{3}}{4}(\ket{3,1}-\ket{3,-1})\\  \ket{y^3} & =  \tfrac{\sqrt{5}}{4i}(\ket{3,3}+\ket{3,-3})+\tfrac{\sqrt{3}}{4i}(\ket{3,1}+\ket{3,-1}) \\  \ket{z^3} & =  \ket{3,0} \end{aligned}$  && \\[22pt] 
 & $ T_{2g,u}$ & $\begin{aligned} \ket{z(x^2-y^2)} & =  \tfrac{1}{\sqrt{2}}(\ket{3,-2}+\ket{3,2})\\  \ket{x(y^2-z^2)} & =  \tfrac{\sqrt{3}}{4}(-\ket{3,3}+\ket{3,-3})-\tfrac{\sqrt{5}}{4}(\ket{3,1}-\ket{3,-1}) \\  \ket{y(z^2-x^2)} & =  \tfrac{\sqrt{3}}{4i}(\ket{3,3}+\ket{3,-3})-\tfrac{\sqrt{5}}{4i}(\ket{3,1}+\ket{3,-1}) \end{aligned}$& &\\[22pt] 
  \hline
$4$ & $A_{1g,u} $ &  $ \ket{A_1}=\sqrt{\tfrac{5}{24}}\ket{4,4}  + \sqrt{\tfrac{7}{12}}\ket{4,0}+\sqrt{\tfrac{5}{24}} \ket{4,-4}$  &  $\ket{\bdelta_{O}}_4$ &  Eq. \eqref{eq:octa4} \\ [5pt] 
 & $E_{g,u}$ &  $\begin{aligned} & \ket{ E,\theta_1}  =  \sqrt{\tfrac{7}{24}}\ket{4,4}  - \sqrt{\tfrac{5}{12}}\ket{4,0}+\sqrt{\tfrac{7}{24}} \ket{4,-4}  \\   &  \ket{ E,\theta_2}  =  \tfrac{1}{\sqrt{2}}(\ket{4,2}+\ket{4,-2})  \end{aligned}$  &$\begin{aligned} & \ket{\bdelta_{D_4}} = \ket{E,\theta_2} \\ &\ket{\bdelta_{T}} = \ket{E,\theta_1}+i\ket{ E,\theta_2}  \end{aligned}$ & $\begin{aligned} & \text{Eq. \eqref{eq:D44}} \\ & \text{Eq. \eqref{eq:tetra4}}  \end{aligned}$\\ [18pt] 
 & $T_{1g,u} $ & $\begin{aligned} \ket{T_1,\xi_1} & = \tfrac{1}{i\sqrt{2}}(\ket{4,4}-\ket{4,-4})\\  \ket{T_1,\xi_2} & =  \tfrac{1}{4i}(\ket{4,3}+\ket{4,-3})+\tfrac{\sqrt{7}}{4i}(\ket{4,1}+\ket{4,-1}) \\  \ket{T_1,\xi_3} & =   \tfrac{1}{4}(-\ket{4,3}+\ket{4,-3})+\tfrac{\sqrt{7}}{4}(\ket{4,1}-\ket{4,-1})  \end{aligned}$  &  & \\ [22pt] 
 & $T_{2g,u} $ & $\begin{aligned} \ket{T_2,\zeta_1} & = -\tfrac{\sqrt{7}}{4i}(\ket{4,3}+\ket{4,-3})+\tfrac{1}{4i}(\ket{4,1}+\ket{4,-1}) \\  \ket{T_2,\zeta_2} & =  \tfrac{\sqrt{7}}{4}(\ket{4,3}-\ket{4,-3})+\tfrac{1}{4}(\ket{4,1}-\ket{4,-1}) \\  \ket{T_2,\zeta_3} & =  \tfrac{1}{i\sqrt{2}}(\ket{4,2}-\ket{4,-2}) \end{aligned}$  &  & \\ [2pt] 
\end{tabular}
\end{ruledtabular}
 \caption{{\bf Splitting of pairing channels in cubic systems.} In the presence of cubic crystal anisotropy, the isotropic pairing channels labeled by angular momentum $J$ are split into cubic pairing channels labeled by cubic representations, as explained in Sec. \ref{sec:pairingclass}. This Table lists the splitting of the isotropic pairing components given in Eq. \eqref{eq:pairJMJ} into pairing functions transforming as partners of the cubic representations; this is shown in the third column, using the notation of Eq. \eqref{eq:spinstate}. We list the splitting of angular momentum channels $J=1,2,3,4$. Note that some cubic representations appear multiple times; the corresponding pairing functions are ``degenerate'' in cubic symmetry. Note further that the parity $g,u$ depends on the quantum numbers $(L,S)$. A number of pairing states which can arise in cubic systems (and thus necessarily have discrete symmetry, see \ref{ssec:cubicsymmetry}) are discussed in Sec. \ref{ssec:gapdiscrete}; these are listed in the fourth column. Column five refers to the specific equation.
 }
\label{tab:cubicsymmetry}
\end{table*}



\end{document}